\documentclass[10 pt]{article}
\usepackage[cp1251]{inputenc}
\usepackage[ukrainian]{babel}
\usepackage{graphicx}
\usepackage{amssymb}

\date{}

\title{Internal energy of many-boson system with three- and four-particle direct correlations taken into account}
\author{I. O. Vakarchuk, O. I. Hryhorchak\\
{\small Department for Theoretical Physics, Ivan Franko National
University of Lviv,}\\
{\small 12, Drahomanov Str., Lviv, UA--79005,
Ukraine}}

\begin{document}
\renewcommand{\abstractname}{Abstract}
\maketitle

\begin{abstract}
In this paper we calculate kinetic, potential and full energy with three- and
four-particle direct correlations taken into account at wide temperature region
 on the base of the density matrix of the interacting Bose-particles [I.
O. Vakarchuk, O. I. Hryhorchak, Journ. Phys. Stud. {\bf 3}, 3005
(2009)]. In the low temperature limit the obtained
expression for the full energy is equal to the wellknown
expression for ground state energy in the approximation of ``two
sums over the wave vector''. The
results of this work can be applied for the numeric calculation of
the heat capacity  of liquid $^4$He
in order to check the theoretical
 and experimental results quantatively, especially in the $\lambda$-transition
region.
\end{abstract}

\section{Вступ}

      Розрахунок внутрішньої енергії такої багатобозонної системи, як рідкий $^4$He має доволі давню історію.
 Великою мірою це пов'язано із прагненням теоретично описати $\lambda$-подібний хід кривої теплоємності в околі точки 
 фазового переходу. Перші кроки були зроблені в напрямі наближеного обчислення енергії основного стану і енергетичного спектру цієї системи
 ще Боголюбовим \cite{B1947} майже сімдесят років тому.
Пізніше він разом із Зубарєвим  у роботі \cite{BZ1955} знайшли хвильову функцію основного стану,
а також хвильові функції нижніх збуджених рівнів для слабонеідеального бозе-газу з допомогою методу колективних змінних.
  
  В границі зникаюче малої взаємодії енергія основного стану бозе-системи була вперше коректно знайдена в роботі
     \cite{BS1957}. Розрахунок проводився з використанням псевдотенціалу. 
  
  Дослідження термодинамічних функцій основного стану рідкого $^4$He, зокрема внутрішньої енергії, 
в наближенні вищому, ніж наближення Боголюбова, було проведено в ряді робіт 
\cite{VakUhn79_VHU79,V1985_89_90,V1988_VH,Hlushak}. Завдяки цим 
роботам було показано, що врахування три- та чотиричастинкових кореляцій покращує значення для енергії основного стану.

   Трохи згодом в  роботах
\cite{Vak_Bab_Rov,Vak_Rov12} з допомогою двочасових
температурних функцій Гріна були отримані вирази для
термодинамічних функцій рідкого $^4$He в широкотемпературному
діапазоні. 

В підході колективних змінних термодинамічні функції багатобозонної системи  в широкотемпературній області
були знайдені в роботі \cite{VPR2007}. Розрахунок проводився
  за  допомогою уcереднення  з матрицею густини взаємодіючих бозе-частинок в наближенні парних кореляцій. 
 Узгодження з експериментальними даними отриманих результатів для  кінетичної, потенціальної та повної енергії 
 є досить добрим \cite{expe1,expe2,expe3,expe4}, однак неповним. Це великою мірою пов'язано з тим, що для матриці густини
 було взято лише наближення парних кореляції. Для більш точних результатів потрібно
 врахувати три- та чотиричастинкові кореляції. Як відомо \cite{Temperly,Krokston}, 
 їх внесок в значення термодинамічних функцій може виявитися досить значним. 
 Врахування внесків три- і чотиричастинкових кореляцій у внутрішню енергію багатобозонної системи і 
 є предметом цієї роботи.

  Вивчення термодинамічних властивостей  $^4$He сьогодні проводиться не тільки для рідкого\cite{SBBLD,LRCG,MaEd}
чи газоподібного стану \cite{Mosameh},
багато робіт присвячені і твердому стану \cite{Chan,HRV,Kim,CaMi,KrCh}. 
Активно вивчається термодинаміка плівок \cite{ADM,CCK,DKFG} та сумішей \cite{ChBr,FoMo}, а також поведінка 
$^4$He в різних середовищах \cite{GoBo,SDG,LCS}.

  Варто також зауважити, що важливе значення у дослідженні
властивостей рідкого $^4$He та  вивченні його термодинамічних функцій
відіграють чисельні методи розрахунку, зокрема такі як: дифузійний метод
Монте Карло (дослідження основного стану) \cite{Vitiello,BMK,BGC}, метод Монте-Карло з використанням
інтегралів за траекторіями \cite{Ceperley,Robo,BSSC} чи функцій Гріна \cite{GFMC},  ``незміщений'' метод Монте-Карло
(unbiased Monte Carlo)
\cite{MoBo2}. При застосуванні згаданих чисельних методів постає проблема задання
міжатомного потенціалу взаємодії в гелії. Для цього в різний час 
використовували потенціал Слетера-Кірквуда\cite{SL_KR}, Ленарда-Джонса \cite{EKZ,MiSc}, Азіза-Сламана
\cite{AZ_SL1,AZ_SL2}. Застосовують також потенціали, які
враховують багаточастинкові кореляції, зокрема тричастинкові
\cite{Br_Cs}. Корисним також є потенціал Аксильрода-Теллера \cite{Ak_Tl}, який
враховує диполь-диполь-дипольні взаємодії та аналітичний потенціал
Паріша-Дикстри \cite{Pr_Dk}. Інший підхід до цієї проблеми був запропований 
у роботах \cite{Vak_Bab_Rov,Rov_dys}, де міжатомний потенціал взаємодії був відновлений 
за експериментальними даними. Результатами саме цього підходу ми скористаємося 
для проведення чисельних розрахунків в цій роботі.

 В наших попередніх статтях \cite{VakHryh1,VakHryh2,VakHryh3} були знайдені матриця густини, 
 статистична сума, дво-, три- і чотиричастинкові структурні фактори
 в широкому інтервалі температур із
врахуванням прямих три- та чотиричастинкових  кореляцій. Отримані результати
стали основою розрахунку виразів для кінетичної,
потенціальної і повної енергія багатобозонної системи 
в наближенні ``двох сум ха хвильовим вектором''.
Чисельний розрахунок термодинамічних функцій  
проводився з урахуванням раніше знайденої ефективної маси \cite{HP2014}.
Отриманий вираз для повної енергії в границі
низьких температур  узгоджується з уже відомим результатом
\cite{VakUhn79_VHU79}. 

\section{Енергiя, розрахована на основi матрицi густини взаємодiючих бозе-частинок без видiлення матрицi 
густини iдеального бозе-газу}
В роботі \cite{VakHryh2} була знайдена статистична
сума багатобозонної системи в наближенні ``двох сум за хвильовим
вектором'' із врахуванням прямих три- та чотиричастинкових
кореляцій на основі матриці густини взаємодіючих бозе-частинок без
віділення матриці густини ідеального бозе-газу у наступному
вигляді: 
\begin{eqnarray}\label{Z_res}
Z&=&e^{-\beta
F_0}\exp\left[\frac{\overline{C}_0}{N}\right]\exp\left[2\sum_{\mathbf{q}_1\neq0}
\frac{\overline{C}_2(\mathbf{q}_1)}{\alpha_{q_1}\tanh\left[\frac{\beta}{2}E_{q_1}\right]}+\frac{2}{N}\sum_{\mathbf{q}_1\neq0}\sum_{\mathbf{q}_2\neq0}\frac{\overline{C}_4(\mathbf{q}_1,\mathbf{q}_2)}
{\alpha_{q_1}\alpha_{q_2}\tanh\left[\frac{\beta}{2}E_{q_1}\right]\tanh\left[\frac{\beta}{2}E_{q_2}\right]}
\right.\nonumber\\
&+&\left.\frac{12}{N}\mathop{\sum_{\mathbf{q}_1\neq0}
\sum_{\mathbf{q}_2\neq0}\sum_{\mathbf{q}_3\neq0}}
\limits_{{\mathbf q}_1+{\mathbf q}_2+{\mathbf q}_3=0}
\frac{{\overline{C}_3}^2({\mathbf{q}}_1,{\mathbf{q}}_2,
{\mathbf{q}}_3)}{\alpha_{q_1}\alpha_{q_2}\alpha_{q_3}
\tanh\left[\frac{\beta}{2}E_{q_1}\right]\tanh\left[\frac{\beta}{2}E_{q_2}\right]\tanh\left[\frac{\beta}{2}E_{q_3}\right]}\right],
\end{eqnarray}
 
де
\begin{eqnarray}
\alpha_q=\sqrt{1+{\frac{2N}{V}\nu_q}\left\slash\frac{\hbar^2q^2}{2m}\right.},\nonumber\\
E_q=\varepsilon_q\alpha_q=\frac{\hbar^2q^2}{2m}\alpha_q,
\end{eqnarray}
\begin{eqnarray}
F_0&=&\frac{N(N-1)}{2V}\nu_0-\sum_{{\mathbf
q}\neq0}\frac{\hbar^2q^2}{8m}(\alpha_q-1)^2
+\frac{1}{\beta}\sum_{{\mathbf q}\neq0}\ln\left(1-e^{-\beta
E_q}\right),
\end{eqnarray}
 $\nu_q=\int e^{-i\bf{qr}}\Phi(r)d\bf{r}$ --- це коефіцієнт
Фур'є енергії парної взаємодії між частинками,
$\beta=1/T$ --- обернена температура. Явні вирази для величин $\overline{C}_2({\bf q_1})$,
$\overline{C}_3({\bf q_1},{\bf q_2},{\bf q_3})$,
$\overline{C}_4({\bf q_1},{\bf q_2})$ наведені у роботі \cite{VakHryh1,VakHryh3}.

 Якщо скористатися виразом (\ref{Z_res}) для статистичної суми  і згідно з формулою
\begin{eqnarray}
E=-\frac{\partial}{\partial\beta}\ln Z
\end{eqnarray}
знайти внутрішню енергію багатобозонної системи в наближенні ``двох сум за хвильовим вектором'', 
то виявиться, що в границі низьких температур вона поводиться правильно і переходить 
у вже відомий \cite{VakUhn79_VHU79} вираз, натомість в границі високих температур дає некоректні результати. 
Це пов'язано з тим, що і статистична сума, якою ми скористалися для розрахунку,
не переходить у класичний вираз при високих температурах, 
як це було показано раніше \cite{VakHryh2}. 
  Для того, щоб отримати коректний вираз для енергії, який би давав правильні результати як в границі низьких, так і в
границі високих температур, потрібно розрахунок енергії проводити з матрицею густини взаємодіючих бозе-частинок із виділеною
матрицею густини ідеального бозе-газу.

\section{Енергiя, розрахована на основi матрицi густини взаємодiючих 
бозе-частинок з видiленою матрицею густини iдеального бозе-газу}

Середню енергію системи $N$ безспінових бозе-частинок з
координатами $({\mathbf r}_1,\ldots,{\mathbf r}_N)$ можна записати
наступним чином:
\begin{eqnarray}
E=\langle \hat{H}\rangle=\langle \hat{K}\rangle+\langle
\hat{\Phi}\rangle,
\end{eqnarray}
де $\hat{H}$ --- гамільтоніан вказаної системи, $\hat{K}$ --- оператор
кінетичної, а $\hat{\Phi}$ --- потенціальної енергії:
\begin{eqnarray}
\hat{K}=-\frac{\hbar^2}{2m}\sum_{j=1}^N \nabla_j^2,
\end{eqnarray}
\begin{eqnarray}\label{Pe}
 \hat{ \Phi}=\sum_{1\leq i<j\leq
N}\Phi(|{\mathbf r}_i-{\mathbf r}_j|),
\end{eqnarray}
Усереднення проводиться з матрицею густини взаємодіючих
бозе-частинок із врахуванням прямих три- і чотиричастинкових
кореляцій із виділеною матрицею густини ідеального бозе-газу:
\begin{eqnarray}
R(\rho|\rho')&=&R_N^0(r|r')P_{pair}(\rho|\rho')P(\rho|\rho')=R_N^0(r|r')\exp[U],
\end{eqnarray}
де $R_N^0(r|r')$ --- матриця густини невзаємодіючих бозе-частинок,
$P_{pair}(\rho|\rho')$ --- фактор, який враховує парні кореляції, а
$P(\rho|\rho')$ --- фактор, який враховує прямі три- і
чотиричастинкові кореляції.
\begin{eqnarray}
R_N^0(r'|r)\!&=&\!\frac{1}{N!}\!\left(\!\frac{m}{2\pi\beta\hbar^2}\!\right)^{\!\!\!3N/2}
 \!\!\!\!\!\!\!\!\sum\limits_Q\!
 \exp\!\left[\!-\frac{m}{2\beta\hbar^2}\sum\limits_{j=1}^N(r_j'\!-\!r_{Q_j})^2\!\right],\nonumber
\end{eqnarray}
де підсумовування за $Q$ означає підсумовування  за всіма перестановками координат частинок.
 Вираз для величини U має наступний вигляд:
  \begin{eqnarray}
U&=&b_0+\sum_{{\mathbf q}\neq0} b_1({\mathbf q})\rho_{{\mathbf
q}}\rho_{-{\mathbf q}_1}'+\sum_{{\mathbf q}\neq0} b_2({\mathbf
q})(\rho_{{\mathbf q}}\rho_{-{\mathbf q}}+\rho_{{\mathbf
q}}'\rho_{-{\mathbf q}}')+c_0+\sum_{{\mathbf q}_1\neq0}
\sum_{i_1=0}^1 \sum_{j_1=0}^1 c_2(1^{j1},-1^{i_1})\rho_{{\mathbf
q}_1}^{j_1}\rho_{-{\mathbf q}_1}^{i_1}\nonumber\\
&+&\frac{1}{\sqrt{N}}\mathop{\sum_{{\mathbf
q}_1\neq0}\sum_{{\mathbf q}_2\neq0}\sum_{{\mathbf q}_3\neq0}}
\limits_{{\mathbf q}_1+{\mathbf q}_2+{\mathbf
q}_3=0}\sum_{i_1,i_2,i_3=0}^1 c_3(1^{i_1},2^{i_2},3^{i_3})\rho_
{{\mathbf q}_1}^{i_1}\rho_{{\mathbf q}_2}^{i_2}\rho_{{\mathbf q}_3}^{i_3}\nonumber\\
&+&\frac{1}{N}\sum_{{\mathbf q}_1\neq0}\sum_{{\mathbf q}_2\neq0}
\sum_{i_1,i_2=0}^1 \sum_{j_1,j_2=0}^1
c_4(1^{j_1},-1^{i_1},2^{j_2},-2^{i_2}) \rho_{{\mathbf
q}_1}^{j_1}\rho_{-{\mathbf q}_1}^{i_1} \rho_{{\mathbf
q}_2}^{j_2}\rho_{-{\mathbf q}_2}^{i_2},
\end{eqnarray}
  де індекси $i_1,i_2,i_3,j_1,j_2$ пробігають значення
$0,1$. Значення $1$ відповідає присутності штриха біля відповідної
величини, а значення $0$ --- відсутності;
\begin{eqnarray}
b_0&=&-\beta E_0+\frac{1}{2}\sum_{{\mathbf
q}\neq0}\ln\left[\frac{\alpha_{q}\tanh\left(\frac{\beta
E_{q}}{2}\right)}{\tanh\left(\frac{\beta
\varepsilon_{q}}{2}\right)}\right]+
\sum_{{\mathbf
q}\neq0}\ln\left(\frac{1-e^{-\beta\varepsilon_{q}}}{1-e^{-\beta
E_{q}}}\right),\\
b_1{( q)}&=&\frac{1}{2} \left[\frac{\alpha_{q}}{\sh(\beta
E_{q})}-\frac{1}{\sh(\beta
\varepsilon_{q})}\right],\\
b_2(q)&=&-\frac{1}{4} \left[\alpha_{q}\cth(\beta
E_{q})-\cth(\beta\varepsilon_{q})\right].
\end{eqnarray}
Явні вирази для величин $c_0, c_2,
c_3, c_4$ є доволі громіздкими і наведені у
роботі \cite{VakHryh3}. 

\section{Середня кiнетична енергiя}
Скористаємося результатами роботи \cite{VPR2007} і запишемо вираз
для кінетичної енергії в наступний спосіб:
\begin{eqnarray}
\left\langle\hat{K}\right\rangle=\left\langle\hat{K_1}\right\rangle+\left\langle\hat{K_2}\right\rangle+\left\langle\hat{K_3}\right\rangle,
\end{eqnarray}
де
  
\begin{eqnarray}
\left\langle\hat{K_1}\right\rangle=\frac{\int d{\bf r}_1...\int
d{\bf
r}_N\left[P_{pair}(\rho|\rho')P(\rho|\rho')\sum\limits_{j=1}^N\left(-\frac{\hbar^2\nabla_j^2}{2m}\right)R_N^0({\bf
r}_1',...,{\bf r}_N'|{\bf r}_1,...,{\bf r}_N)\right]_{{\bf
r}_1'={\bf r}_1,...,{\bf r}_N'={\bf r}_N}}{\int d{\bf r}_1...\int
d{\bf r}_N P_{pair}(\rho|\rho)P(\rho|\rho)R_N^0({\bf r}_1,...,{\bf
r}_N|{\bf r}_1,...,{\bf r}_N)}.
\end{eqnarray}
Застосувавши теорму Блоха: $-\partial
R_N^0/\partial\beta=\hat{K}R_N^0$ до написаного вище виразу,
знайдемо, що середнє
\begin{eqnarray}
\left\langle\hat{K_1}\right\rangle=\frac{\partial}{\partial\beta'}\ln\left\{\int
d{\bf r}_1...\int d{\bf r}_N
P_{pair}(\rho|\rho)P(\rho|\rho)R_N^0({\bf r}_1,...,{\bf r}_N|{\bf
r}_1,...,{\bf r}_N;\beta')\right\}_{\beta'=\beta},
\end{eqnarray}
яке у прийнятому нами наближенні ``двох сум за хвильовим вектором''
можна подати у вигляді:
\begin{eqnarray}
\left\langle\hat{K_1}\right\rangle=\frac{\partial}{\partial\beta'}\ln\left\{\int
d{\bf r}_1...\int d{\bf r}_N P_{pair}(\rho|\rho)R_N^0({\bf
r}_1,...,{\bf r}_N|{\bf r}_1,...,{\bf
r}_N;\beta')\right\}_{\beta'=\beta}+\frac{\partial}{\partial\beta'}\ln\left[\left\langle
P(\rho|\rho)\right\rangle\right]_{\beta'=\beta}.
\end{eqnarray}
Перший доданок у написаному виразі був знайдений в роботі
\cite{VPR2007}, а середнє $\left\langle P(\rho|\rho)\right\rangle$
має такий зміст:
\begin{eqnarray}
\left\langle P(\rho|\rho)\right\rangle=\frac{\int d{\bf
r}_1...\int d{\bf r}_N R_N^0({\bf r}_1,...,{\bf r}_N|{\bf
r}_1,...,{\bf r}_N)P_{pair}(\rho|\rho)P(\rho|\rho)}{\int d{\bf
r}_1...\int d{\bf r}_N R_N^0({\bf r}_1,...,{\bf r}_N|{\bf
r}_1,...,{\bf r}_N)P_{pair}(\rho|\rho)}.
\end{eqnarray}
Це середнє вже було знайдене в роботі \cite{VakHryh3}:
\begin{eqnarray}\label{P_ser}
\left\langle
P(\rho|\rho)\right\rangle&=&\exp\left\{C_0+2\sum_{\mathbf{q}_1\neq0}C_2({\mathbf
q}_1)\frac{S_0(
q_1)}{1+\lambda_{q_1}S_0(q_1)}+\frac{2}{N}\sum_{\mathbf{q}_1\neq0}\sum_{\mathbf{q}_2\neq0}C_4({\mathbf
q}_1,{\mathbf q}_2)\frac{S_0(
q_1)}{1+\lambda_{q_1}S_0(q_1)}\frac{S_0(
q_2)}{1+\lambda_{q_2}S_0(q_2)}\right.\nonumber\\
&+&\frac{2}{N}\mathop{\sum_{\mathbf{q}_1
\neq0}\sum_{\mathbf{q}_2\neq0}\sum_{\mathbf{q}_3\neq0}}\limits_{\mathbf{q}_1+\mathbf{q}_2+\mathbf{q}_3=0}
C_3({\mathbf q}_1,{\mathbf q}_2,{\mathbf q}_3)\frac{S_0^{(3)}(
q_1,q_2,q_3)}{[1+\lambda_{q_1}S_0(q_1)][1+\lambda_{q_2}S_0(q_2)][1+\lambda_{q_3}S_0(q_3)]}\nonumber\\
&+&\left.\frac{12}{N}\mathop{\sum_{\mathbf{q}_1\neq0}\sum_{\mathbf{q}_2\neq0}\sum_{\mathbf{q}_3\neq0}}\limits_{\mathbf{q}_1+\mathbf{q}_2+\mathbf{q}_3=0}
C_3^2({\mathbf q}_1,{\mathbf q}_2,{\mathbf
q}_3)\frac{S_0(q_1)S_0(q_2)S_0(q_3)}{[1+\lambda_{q_1}S_0(q_1)][1+\lambda_{q_2}S_0(q_2)][1+\lambda_{q_3}S_0(q_3)]}\right\},
\end{eqnarray}
  де від величини $\beta'$ залежать лише парний
$S_0(q_i),i=1,2,3$ і тричастинковий $S_0^{(3)}( q_1,q_2,q_3)$
структурні фактори ідеального бозе-газу.

В результаті для величини $\left\langle\hat{K_1}\right\rangle$
отримаємо такий вираз:

  \begin{eqnarray}\label{K1_res}
\left\langle\hat{K_1}\right\rangle&=&\sum_{{\mathbf
q}\neq0}\frac{\varepsilon_q}{z_0^{-1}e^{\beta\varepsilon_q}-1}+\frac{1}{2}\sum_{{\mathbf
q}\neq0}\frac{\lambda_{q}}{1+\lambda_{q}S_0(q)}\frac{\partial
S_0(q)}{\partial\beta}-\frac{1}{8N}\sum_{\mathbf{q}_1\neq0}\sum_{\mathbf{q}_2\neq0}
\left[\prod_{i=1}^2\frac{\lambda_{q_i}}{1+\lambda_{q_i}S_0(q_i)}\right]\nonumber\\
&\times&\left\{\frac{\partial S_0^{(4)}({\bf q}_1,-{\bf q}_1, {\bf
q}_2,-{\bf
q}_2)}{\partial\beta}-\left[\sum_{j=1}^2\frac{\lambda_{q_j}}{1+\lambda_{q_j}S_0(q_j)}\frac{\partial
S_0(q)}{\partial\beta}\right]S_0^{(4)}({\bf q}_1,-{\bf q}_1, {\bf
q}_2,-{\bf q}_2)\right\}\nonumber\\
&+&\frac{1}{3!}\frac{1}{2N}\mathop{\sum_{\mathbf{q}_1\neq0}
\sum_{\mathbf{q}_2\neq0}\sum_{\mathbf{q}_3\neq0}}
\limits_{{\mathbf q}_1+{\mathbf q}_2+{\mathbf
q}_3=0}\left[\prod_{i=1}^3\frac{\lambda_{q_i}}{1+\lambda_{q_i}S_0(q_i)}\right]
\left\{2S_0^{(3)}({\bf q}_1,{\bf q}_2, {\bf q}_3)\frac{\partial
S_0^{(3)}({\bf q}_1,{\bf q}_2, {\bf
q}_3)}{\partial\beta}\right.\nonumber\\
&-&\left.\left[\sum_{j=1}^3\frac{\lambda_{q_j}}{1+\lambda_{q_j}S_0(q_j)}\frac{\partial
S_0(q)}{\partial\beta}\right]\left[S_0^{(3)}({\bf q}_1,{\bf q}_2,
{\bf q}_3)\right]^2\right\}-
2\sum_{\mathbf{q}_1\neq0}\frac{C_2({\mathbf
q}_1)}{[1+\lambda_{q_1}S_0(q_1)]^2}\frac{\partial S_0(
q_1)}{\partial\beta}\nonumber\\
&-&\frac{2}{N}\sum_{\mathbf{q}_1\neq0}\sum_{\mathbf{q}_2\neq0}C_4({\mathbf
q}_1,{\mathbf q}_2)
\mathop{\sum_{\{i,j\}=\{1,2\},}}
\limits_{\{2,3\},\{3,1\}}\frac{1}{[1+\lambda_{q_i}S_0(q_i)]^2}\frac{S_0(q_j)}{1+\lambda_{q_j}S_0(q_j)}\frac{\partial
S_0(
q_i)}{\partial\beta}\nonumber\\
&-&\frac{2}{N}\mathop{\sum_{\mathbf{q}_1
\neq0}\sum_{\mathbf{q}_2\neq0}\sum_{\mathbf{q}_3\neq0}}\limits_{\mathbf{q}_1+\mathbf{q}_2+\mathbf{q}_3=0}
\frac{C_3({\mathbf q}_1,{\mathbf q}_2,{\mathbf
q}_3)}{[1+\lambda_{q_1}S_0(q_1)][1+\lambda_{q_2}S_0(q_2)][1+\lambda_{q_3}S_0(q_3)]}\nonumber\\
&\times&\left(\frac{\partial S_0^{(3)}(
q_1,q_2,q_3)}{\partial\beta}-\left[\sum_{j=1}^3\frac{\lambda_{q_j}}{1+\lambda_{q_j}S_0(q_j)}\frac{\partial
S_0(q)}{\partial\beta}\right]S_0^{(3)}(
q_1,q_2,q_3)\right)\nonumber\\
&+&\frac{12}{N}\mathop{\sum_{\mathbf{q}_1\neq0}\sum_{\mathbf{q}_2\neq0}\sum_{\mathbf{q}_3\neq0}}\limits_{\mathbf{q}_1+\mathbf{q}_2+\mathbf{q}_3=0}
C_3^2({\mathbf q}_1,{\mathbf q}_2,{\mathbf
q}_3)
\mathop{\sum_{\{i,j,l\}=\{1,2,3\},}}
\limits_{\{2,3,1\},\{3,1,2\}}\frac{1}{[1+\lambda_{q_i}S_0(q_i)]^2}\frac{S_0(q_j)}{1+\lambda_{q_j}S_0(q_j)}\frac{S_0(q_l)}{1+\lambda_{q_l}S_0(q_l)}
\frac{\partial
S_0( q_i)}{\partial\beta}.
\end{eqnarray}
 
 Тепер перейдімо до розрахунку наступних середніх
 $\left\langle\hat{K_2}\right\rangle$ і
 $\left\langle\hat{K_3}\right\rangle$, вираз для яких легко
 записати, скориставшись результатами роботи \cite{VPR2007}:
\begin{eqnarray}
\left\langle\hat{K_2}\right\rangle=-\frac{\hbar^2}{2m}\left\langle\sum_{j=1}^N\left(\nabla_j^2U+[\nabla_jU]^2\right)\mid_{r_1'=r_1,...r_N'=r_N}\right\rangle,
\nonumber
\end{eqnarray}
\begin{eqnarray}
\left\langle\hat{K_3}\right\rangle=\frac{\hbar^2}{4m}\left\langle\sum_{j=1}^N(\nabla_j^2U+[\nabla_jU]^2\right\rangle.\nonumber
\end{eqnarray}
Якщо перейти до представлення колективних змінних, то для величин
$\left\langle\hat{K_2}\right\rangle$ і
$\left\langle\hat{K_3}\right\rangle$ отримаємо наступні вирази:
  
\begin{eqnarray}\label{K2}
\left\langle\hat{K_2}\right\rangle&=&\sum_{{\mathbf
k}_1\neq0}\frac{\hbar^2k_1^2}{2m}\left(\left\langle\rho_{{\mathbf
k}_1}\frac{\partial U}{\partial\rho_{{\mathbf
k}_1}}\mid_{\rho=\rho'}\right\rangle-\left\langle\frac{\partial^2
U}{\partial\rho_{{\mathbf k}_1}\partial\rho_{-{\mathbf
k}_1}}\mid_{\rho=\rho'}\right\rangle-\left\langle\frac{\partial
U}{\partial\rho_{{\mathbf k}_1}}\frac{\partial
U}{\partial\rho_{-{\mathbf
k}_1}}\mid_{\rho=\rho'}\right\rangle\right)\nonumber\\
&+&\frac{1}{\sqrt{N}}\mathop{\sum_{{\mathbf
k}_1\neq0}\sum_{{\mathbf k}_2\neq0}}\limits_{{\mathbf
k}_1+{\mathbf k}_2\neq0}\frac{\hbar^2({\mathbf k}_1{\mathbf
k}_2)}{2m}\left(\left\langle\rho_{{\mathbf k}_1+{\mathbf
k}_2}\frac{\partial^2 U}{\partial\rho_{{\mathbf
k}_1}\partial\rho_{{\mathbf
k}_2}}\mid_{\rho=\rho'}\right\rangle+\left\langle\rho_{{\mathbf
k}_1+{\mathbf k}_2}\frac{\partial U}{\partial\rho_{{\mathbf
k}_1}}\frac{\partial U}{\partial\rho_{{\mathbf
k}_2}}\mid_{\rho=\rho'}\right\rangle\right),
\end{eqnarray}
\begin{eqnarray}
\left\langle\hat{K_3}\right\rangle&=&\frac{1}{2}\sum_{{\mathbf
k}_1\neq0}\frac{\hbar^2k_1^2}{2m}\left(-\left\langle\rho_{{\mathbf
k}_1}\frac{\partial U}{\partial\rho_{{\mathbf
k}_1}}\right\rangle+\left\langle\frac{\partial^2
U}{\partial\rho_{{\mathbf k}_1}\partial\rho_{-{\mathbf
k}_1}}\right\rangle+\left\langle\frac{\partial
U}{\partial\rho_{{\mathbf k}_1}}\frac{\partial
U}{\partial\rho_{-{\mathbf
k}_1}}\right\rangle\right)\nonumber\\
&-&\frac{1}{2\sqrt{N}}\mathop{\sum_{{\mathbf
k}_1\neq0}\sum_{{\mathbf k}_2\neq0}}\limits_{{\mathbf
k}_1+{\mathbf k}_2\neq0}\frac{\hbar^2({\mathbf k}_1{\mathbf
k}_2)}{2m}\left(\left\langle\rho_{{\mathbf k}_1+{\mathbf
k}_2}\frac{\partial^2 U}{\partial\rho_{{\mathbf
k}_1}\partial\rho_{{\mathbf
k}_2}}\right\rangle+\left\langle\rho_{{\mathbf k}_1+{\mathbf
k}_2}\frac{\partial U}{\partial\rho_{{\mathbf k}_1}}\frac{\partial
U}{\partial\rho_{{\mathbf k}_2}}\right\rangle\right).
\end{eqnarray}
 
 Знак середнього $\langle...\rangle$ тут знову має зміст
усереднення з матрицею густини взаємодіючих бозе-частинок із
врахуванням прямих три- та чотиричастинкових  кореляцій:
\begin{eqnarray}
\langle...\rangle=\frac{\int d{\mathbf r}_j\ldots\int d{\mathbf
r}_N R_N^0(r|r) P_{pair}(\rho|\rho')P(\rho|\rho)(...)}{\int
d{\mathbf r}_j\ldots\int d{\mathbf r}_N R_N^0(r|r')
P_{pair}(\rho|\rho)P(\rho|\rho)}.\nonumber
\end{eqnarray}

 Зауважимо також, що
\begin{eqnarray}\label{dUdrho}
\left. \frac{\partial U}{\partial \rho_{{\mathbf
k}_1}}\right|_{\rho=\rho'}&=&-\frac{\lambda_{k_1}}{2}\rho_{-{\mathbf
k}_1}+4C_2({\mathbf k}_1)\rho_{-{\mathbf
k}_1}
+\frac{6}{\sqrt{N}}\mathop{\sum_{\mathbf{q}_2\neq0}\sum_{\mathbf{q}_3\neq0}}\limits_{\mathbf{k}_1+\mathbf{q}_2+\mathbf{q}_3=0}
C_3({\mathbf k}_1,{\mathbf q}_2,{\mathbf q}_3)\rho_{{\mathbf
q}_2}\rho_{{\mathbf
q}_3}\nonumber\\
&+&\frac{8}{N}\sum_{\mathbf{q}_1\neq0}\sum_{\mathbf{q}_2\neq0}C_4({\mathbf
k}_1,{\mathbf q}_2)\rho_{-{\mathbf k}_1}\rho_{{\mathbf
q}_2}\rho_{-{\mathbf q}_2},
\end{eqnarray}
\begin{eqnarray}\label{d2Udrho}
\left.\frac{\partial^2 U}{\partial \rho_{{\mathbf
k}_1}\rho_{-{\mathbf
k}_1}}\right|_{\rho=\rho'}&=&2b_2(k_1)+\widetilde{C}_2({\mathbf
k}_1)
+\frac{4}{N}\sum_{\mathbf{q}_2\neq0}\widetilde{C}_4({\mathbf
k}_1,{\mathbf q}_2)\rho_{{\mathbf q}_2}\rho_{-{\mathbf q}_2},
\end{eqnarray}
\begin{eqnarray}\label{d2Udrho2}
\left.\frac{\partial^2 U}{\partial \rho_{{\mathbf
k}_1}\rho_{{\mathbf
k}_2}}\right|_{\rho=\rho'}&=&\frac{6}{\sqrt{N}}\sum_{\mathbf{q}_3\neq0}\widetilde{C}_3({\mathbf
k}_1,{\mathbf k}_2,{\mathbf q}_3)\rho_{{\mathbf q}_3}
+\frac{4}{N}\widetilde{C}_4'({\mathbf
k}_1,{\mathbf k}_2)\rho_{-{\mathbf k}_1}\rho_{-{\mathbf k}_2},
\end{eqnarray}
де
\begin{eqnarray}
&&\widetilde{C}_2({\mathbf k}_1)=c_2(1,-1)+c_2(-1,1),\nonumber\\
&&\widetilde{C}_4({\mathbf k}_1,{\mathbf
q}_2)=\sum_{j_2,i_2=0}^1c_4(1,-1,2^{j_2},-2^{i_2}),\nonumber\\
&&\widetilde{C}_3({\mathbf k}_1,{\mathbf k}_2,{\mathbf
q}_3)=\sum_{i_3=0}^1c_3(1,2,3^{i_3}),\nonumber
\end{eqnarray}
\begin{eqnarray}
\widetilde{C}_4'({\mathbf k}_1,{\mathbf k}_2)
=\sum_{j_1,i_2=0}^1[c_4(1,-1^{i_1},2,-2^{i_2})+c_4(-1^{i_1},1,2,-2^{-i_2})+c_4(1,-1^{i_1},-2^{j_2},2)
+c_4(-1^{i_1},1,-2^{i_2},2)].
\end{eqnarray}

 Явні вирази для величин $\widetilde{C}_2({\mathbf
k}_1)$, $\widetilde{C}_3({\mathbf k}_1,{\mathbf k}_2,{\mathbf
q}_3)$, $\widetilde{C}_4({\mathbf k}_1,{\mathbf q}_2)$ наведені в Додатку 1.
Величини  $C_2({\mathbf
k}_1)$, $C_3({\mathbf k}_1,{\mathbf k}_2,{\mathbf
q}_3)$, $C_4({\mathbf k}_1,{\mathbf q}_2)$ були знайдені раніше \cite{VakHryh3}. 

Підставляючи у рівність (\ref{K2}) отримані вирази
(\ref{dUdrho}), (\ref{d2Udrho}), (\ref{d2Udrho2}), в прийнятому
нами наближенні ``двох сум за хвильовим вектором'' знайдемо:
  
\begin{eqnarray}\label{K2_res}
\left\langle\hat{K_2}\right\rangle&=&\sum_{{\mathbf
k}_1\neq0}\frac{\hbar^2k_1^2}{2m}\left\{-\frac{\lambda_{k_1}}{4}(\lambda_{k_1}+2)\langle\rho_{{\mathbf
k}_1}\rho_{-{\mathbf k}_1}\rangle+2C_2({\mathbf
k}_1)(\lambda_{k_1}+1)\langle\rho_{{\mathbf k}_1}\rho_{-{\mathbf
k}_1}\rangle-\widetilde{C}_2({\mathbf k}_1)-2b_2(\bf{k}_1)\right.\nonumber\\
&+&\frac{3(\lambda_{k_1}+1)}{\sqrt{N}}\mathop{\sum_{\mathbf{q}_2\neq0}\sum_{\mathbf{q}_3\neq0}}\limits_{\mathbf{k}_1+\mathbf{q}_2+\mathbf{q}_3=0}
C_3({\mathbf k}_1,{\mathbf q}_2,{\mathbf
q}_3)\langle\rho_{{\mathbf k}_1}\rho_{{\mathbf q}_2}\rho_{{\mathbf
q}_3}\rangle+\frac{4(\lambda_{k_1}+1)}{N}\sum_{\mathbf{q}_2\neq0}C_4({\mathbf
k}_1,{\mathbf q}_2)\langle\rho_{{\mathbf k}_1}\rho_{-{\mathbf
k}_1}\rho_{{\mathbf q}_2}\rho_{-{\mathbf q}_2}\rangle\nonumber\\
&-&\frac{18}{N}\mathop{\sum_{\mathbf{q}_2\neq0}\sum_{\mathbf{q}_3\neq0}}\limits_{\mathbf{k}_1+\mathbf{q}_2+\mathbf{q}_3=0}
C_3^2({\mathbf k}_1,{\mathbf q}_2,{\mathbf
q}_3)\langle\rho_{-{\mathbf q}_2}\rho_{-{\mathbf
q}_3}\rho_{{\mathbf q}_2}\rho_{{\mathbf
q}_3}\rangle\left.-
\frac{4}{N}\sum_{{\mathbf q}_2\neq0}\widetilde{C}_4({\mathbf
k}_1,{\mathbf q}_2)\langle\rho_{{\mathbf q}_2}\rho_{-{\mathbf
q}_2}\rangle\right\}\nonumber\\
&+&\frac{1}{\sqrt{N}}\mathop{\sum_{{\mathbf
k}_1\neq0}\sum_{{\mathbf k}_2\neq0}}\limits_{{\mathbf
k}_1+{\mathbf k}_2\neq0}\frac{\hbar^2({\mathbf k}_1{\mathbf
k}_2)}{2m}\left\{\frac{6}{\sqrt{N}}\widetilde{C}_3({\mathbf
k}_1,{\mathbf k}_2,-{\mathbf k}_1-{\mathbf
k}_2)\langle\rho_{{\mathbf k}_1+{\mathbf k}_2}\rho_{-{\mathbf
k}_1-{\mathbf
k}_2}\rangle+\frac{\lambda_{k_1}\lambda_{k_2}}{4}\langle\rho_{{\mathbf
k}_1+{\mathbf k}_2}\rho_{-{\mathbf k}_1}\rho_{-{\mathbf
k}_2}\rangle\right.\nonumber\\
&-&\left.\frac{3}{\sqrt{N}}C_3({\mathbf
k}_1,{\mathbf k}_2,-{\mathbf k}_1-{\mathbf
k}_2)\left(\lambda_{k_1}\langle\rho_{{\mathbf k}_1+{\mathbf k}_2}\rho_{-{\mathbf
k}_1-{\mathbf k}_1}\rho_{{\mathbf k}_1}\rho_{-{\mathbf
k}_2}\rangle+\lambda_{k_2}\langle\rho_{{\mathbf k}_1+{\mathbf k}_2}\rho_{-{\mathbf
k}_1-{\mathbf k}_2}\rho_{{\mathbf k}_2}\rho_{-{\mathbf
k}_2}\rangle\right)\right\}.
\end{eqnarray}
Аналогічно, для величини $\left\langle\hat{K_3}\right\rangle$
матимемо:
\begin{eqnarray}\label{K3_res}
\left\langle\hat{K_3}\right\rangle&=&\sum_{{\mathbf
k}_1\neq0}\frac{\hbar^2k_1^2}{2m}\left\{\frac{\lambda_{k_1}}{2}(\lambda_{k_1}+1)\langle\rho_{{\mathbf
k}_1}\rho_{-{\mathbf k}_1}\rangle-2C_2({\mathbf
k}_1)(2\lambda_{k_1}+1)\langle\rho_{{\mathbf k}_1}\rho_{-{\mathbf
k}_1}\rangle-\frac{\lambda_{k_1}}{2}+2C_2({\mathbf k}_1)\right.\nonumber\\
&-&\frac{3(2\lambda_{k_1}+1)}{\sqrt{N}}\mathop{\sum_{\mathbf{q}_2\neq0}\sum_{\mathbf{q}_3\neq0}}\limits_{\mathbf{k}_1+\mathbf{q}_2+\mathbf{q}_3=0}
C_3({\mathbf k}_1,{\mathbf q}_2,{\mathbf
q}_3)\langle\rho_{{\mathbf k}_1}\rho_{{\mathbf q}_2}\rho_{{\mathbf
q}_3}\rangle-\frac{4(2\lambda_{k_1}+1)}{N}\sum_{\mathbf{q}_2\neq0}C_4({\mathbf
k}_1,{\mathbf q}_2)\langle\rho_{{\mathbf k}_1}\rho_{-{\mathbf
k}_1}\rho_{{\mathbf q}_2}\rho_{-{\mathbf q}_2}\rangle\nonumber\\
&+&
\frac{36}{N}\mathop{\sum_{\mathbf{q}_2\neq0}\sum_{\mathbf{q}_3\neq0}}\limits_{\mathbf{k}_1+\mathbf{q}_2+\mathbf{q}_3=0}
C_3^2({\mathbf k}_1,{\mathbf q}_2,{\mathbf
q}_3)\langle\rho_{-{\mathbf q}_2}\rho_{-{\mathbf
q}_3}\rho_{{\mathbf q}_2}\rho_{{\mathbf
q}_3}\rangle\left.+
\frac{4}{N}\sum_{{\mathbf q}_2\neq0}C_4({\mathbf k}_1,{\mathbf
q}_2)\langle\rho_{{\mathbf q}_2}\rho_{-{\mathbf
q}_2}\rangle\right\}\nonumber\\
&-&\frac{1}{\sqrt{N}}\mathop{\sum_{{\mathbf
k}_1\neq0}\sum_{{\mathbf k}_2\neq0}}\limits_{{\mathbf
k}_1+{\mathbf k}_2\neq0}\frac{\hbar^2({\mathbf k}_1{\mathbf
k}_2)}{2m}\left\{\frac{6}{\sqrt{N}}C_3({\mathbf k}_1,{\mathbf
k}_2,-{\mathbf k}_1-{\mathbf k}_2)\langle\rho_{{\mathbf
k}_1+{\mathbf k}_2}\rho_{-{\mathbf k}_1-{\mathbf
k}_2}\rangle+\frac{\lambda_{k_1}\lambda_{k_2}}{2}\langle\rho_{{\mathbf
k}_1+{\mathbf k}_2}\rho_{-{\mathbf k}_1}\rho_{-{\mathbf
k}_2}\rangle\right.\nonumber\\
&-&\left.\frac{6}{\sqrt{N}}C_3({\mathbf
k}_1,{\mathbf k}_2,-{\mathbf k}_1-{\mathbf
k}_2)\left(\lambda_{k_1}\langle\rho_{{\mathbf k}_1+{\mathbf k}_2}\rho_{-{\mathbf
k}_1-{\mathbf k}_2}\rho_{{\mathbf k}_1}\rho_{-{\mathbf
k}_1}\rangle+\lambda_{k_2}\langle\rho_{{\mathbf k}_1+{\mathbf k}_2}\rho_{-{\mathbf
k}_1-{\mathbf k}_2}\rho_{{\mathbf k}_2}\rho_{-{\mathbf
k}_2}\rangle\right)\right\}.
\end{eqnarray}
Підставивши у (\ref{K2_res}) і (\ref{K3_res}) відповідні вирази для всіх середніх $\langle...\rangle$ (вони є 
наведені в роботі \cite{VakHryh3}) 
і обмежившись наближенням ``двох сум за хвильовим вектором'', отримаємо: 
\begin{eqnarray}\label{K23_res}
\left\langle\hat{K_{23}}\right\rangle&=&\left\langle\hat{K_2}\right\rangle+\left\langle\hat{K_3}\right\rangle=
\sum_{{\mathbf
k}_1\neq0}\frac{\hbar^2k_1^2}{2m}\left\{\frac{\lambda_{k_1}^2}{4}S(k_1)-2C_2({\mathbf
k}_1)\lambda_{k_1}\frac{S_0(k_1)}{1+\lambda_{k_1}S_0(k_1)}
+(2C_2({\mathbf k}_1)-\widetilde{C}_2({\mathbf k}_1))\right.\nonumber\\
&-&\frac{3\lambda_{k_1}}{\sqrt{N}}\mathop{\sum_{{\bf q}_2\neq0}\sum_{{\mathbf q}_3\neq0}}
\limits_{{\mathbf k}_1+{\mathbf q}_2+{\mathbf q}_3=0}
C_3({\mathbf k}_1,{\mathbf q}_2,{\mathbf
q}_3)S^{(3)}({\bf k}_1,{\bf q}_2,{\bf q}_3)-\frac{4\lambda_{k_1}}{N}\sum_{{\mathbf q}_2\neq0}C_4({\mathbf
k}_1,{\mathbf q}_2)\frac{S_0(k_1)}{1+\lambda_{k_1}S_0(k_1)}
\frac{S_0(q_2)}{1+\lambda_{q_2}S_0(k_2)}\nonumber\\
&+&\frac{18}{N}\mathop{\sum_{{\mathbf q}_2\neq0}\sum_{{\mathbf q}_3\neq0}}
\limits_{{\mathbf k}_1+{\mathbf q}_2+{\mathbf q}_3=0}
C_3^2({\mathbf k}_1,{\mathbf q}_2,{\mathbf
q}_3)\frac{S_0(q_2)}{1+\lambda_{q_2}S_0(q_2)}\frac{S_0(q_3)}{1+\lambda_{q_3}S_0(q_3)}\left.
+\frac{4}{N}\sum_{{\mathbf q}_2\neq0}(C_4({\mathbf
k}_1,{\mathbf q}_2)-\widetilde{C}_4({\mathbf
k}_1,{\mathbf q}_2))\frac{S_0(q_2)}{1+\lambda_{q_2}S_0(q_2)}\right\}\nonumber\\
&-&\frac{1}{\sqrt{N}}\mathop{\sum_{{\mathbf
k}_1\neq0}\sum_{{\mathbf k}_2\neq0}}\limits_{{\mathbf
k}_1+{\mathbf k}_2\neq0}\frac{\hbar^2({\mathbf k}_1{\mathbf
k}_2)}{2m}\left\{\frac{6}{\sqrt{N}}(C_3({\mathbf
k}_1,{\mathbf k}_2,-{\mathbf k}_1-{\mathbf
k}_2)-\widetilde{C}_3({\mathbf
k}_1,{\mathbf k}_2,-{\mathbf k}_1-{\mathbf
k}_2))\frac{S_0(|\bf{k}_1+\bf{k}_2|)}{1+\lambda_{|\bf{k}_1+\bf{k}_2|}S_0(|\bf{k}_1+\bf{k}_2|)}\right.\nonumber\\
&+&\frac{\lambda_{k_1}\lambda_{k_2}}{4}S^{(3)}({\bf k}_1,{\bf k}_2,-{\bf k}_1-{\bf k}_2)
-\left.\frac{3}{\sqrt{N}}C_3({\mathbf
k}_1,{\mathbf k}_2,-{\mathbf k}_1-{\mathbf
k}_2)\left(\lambda_{k_1}\frac{S_0(k_1)}{1+\lambda_{k_1}S_0(k_1)}
\frac{S_0(|\bf{k}_1+\bf{k}_2|)}{1+\lambda_{|\bf{k}_1+\bf{k}_2|}S_0(|\bf{k}_1+\bf{k}_2|)}\right.\right.\nonumber\\
&+&\left.\left.\lambda_{k_2}\frac{S_0(k_2)}{1+\lambda_{k_2}S_0(k_2)}
\frac{S_0(|\bf{k}_1+\bf{k}_2|)}{1+\lambda_{|\bf{k}_1+\bf{k}_2|}S_0(|\bf{k}_1+\bf{k}_2|)}\right)\right\},
\end{eqnarray}
де $S(k_1)$, $S^{(3)}({\bf k}_1,{\bf q}_2,{\bf q}_3)$ --- це дво- і тричастинковий структурний фактор 
відповідно \cite{VakHryh3}.

\section{Середня потенцiальна енергiя}

 В представленні колективних змінних потенціальна енергія
(\ref{Pe}) запишеться так:
\begin{eqnarray}
\Phi=\frac{N(N-1)}{2V}\nu_0+\frac{N}{2V}\sum_{{\mathbf
q}\neq0}\nu_q (\rho_{{\mathbf q}}\rho_{-{\mathbf q}}-1).
\end{eqnarray}
Беручи до уваги явний вигляд для величини $S(q)=\langle\rho_{{\mathbf q}}\rho_{-{\mathbf q}}\rangle$ 
\cite{VakHryh3}, а також рівність
\begin{eqnarray}
\frac{N}{2V}\nu_q=\frac{\hbar^2}{8m}(\alpha_q^2-1),
\end{eqnarray}
 для середнього значення потенціальної енергії
отримаємо наступний вираз: 
  
\begin{eqnarray}\label{Penergy_res}
\left\langle\Phi\right\rangle&=&\frac{N(N-1)}{2V}\nu_0 +\sum_{{\mathbf
q}\neq0}\frac{\hbar^2}{8m}(\alpha_q^2-1)\left(\frac{S_0(q)}{1+\lambda_{q}S_0(q)}-1\right)\nonumber\\
&-& \frac{1}{2N}\sum_{{\mathbf
q_1}\neq0}\frac{\hbar^2}{8m}(\alpha_{q_1}^2-1)\frac{1}{[1+\lambda_{q_1}S_0(q_1)]^2}\sum_{\mathbf{q}_2\neq0}\frac{\lambda_{q_2}}{1+\lambda_{q_2}S_0(q_2)}
S_0^{(4)}({\bf q}_1,-{\bf q}_1, {\bf
q}_2,-{\bf q}_2)\nonumber\\
&+&\frac{1}{2N}\sum_{{\mathbf
q_1}\neq0}\frac{\hbar^2}{8m}(\alpha_{q_1}^2-1)\frac{1}{[1+\lambda_{q_1}S_0(q_1)]^2}\mathop{\sum_{\mathbf{q}_2\neq0}\sum_{\mathbf{q}_3\neq0}}
\limits_{{\mathbf q}_1+{\mathbf q}_2+{\mathbf
q}_3=0}\frac{\lambda_{q_2}}{1+\lambda_{q_2}S_0(q_2)}\frac{\lambda_{q_3}}{1+\lambda_{q_3}S_0(q_3)}
\left[S_0^{(3)}({\bf q}_1,{\bf q}_2, {\bf
q}_3)\right]^2\nonumber\\
&+&4\sum_{{\mathbf
q_1}\neq0}\frac{\hbar^2}{8m}(\alpha_{q_1}^2-1)C_2({\mathbf{q}_1})\frac{S_0^2(q_1)}{[1+\lambda_{q_1}S_0(q_1)]^2}+
\frac{12}{N}\sum_{{\mathbf
q_1}\neq0}\frac{\hbar^2}{8m}(\alpha_{q_1}^2-1)\frac{S_0(q_1)}{[1+\lambda_{q_1}S_0(q_1)]^2}
\mathop{\sum_{\mathbf{q}_2\neq0}\sum_{\mathbf{q}_3\neq0}}
\limits_{{\mathbf q}_1+{\mathbf q}_2+{\mathbf q}_3=0} C_3({\mathbf
q}_1,{\mathbf q}_2, {\mathbf
q}_3)\nonumber\\
&\times&\frac{S_0^{(3)}({\bf q}_1,{\bf q}_2, {\bf
q}_3)}{[1+\lambda_{q_2}S_0(q_2)][1+\lambda_{q_3}S_0(q_3)]}+\frac{8}{N}\sum_{{\mathbf
q_1}\neq0}\frac{\hbar^2}{8m}(\alpha_{q_1}^2-1)\frac{S_0^2(q_1)}{[1+\lambda_{q_1}S_0(q_1)]^2}
\sum_{\mathbf{q}_2\neq0}C_4({\mathbf q}_1,{\mathbf
q}_2)\frac{S_0(q_2)}{1+\lambda_{q_2}S_0(q_2)}\nonumber\\
&+&\frac{72}{N}\sum_{{\mathbf
q_1}\neq0}\frac{\hbar^2}{8m}(\alpha_{q_1}^2-1)\frac{S_0^2(q_1)}{[1+\lambda_{q_1}S_0(q_1)]^2}
\mathop{\sum_{\mathbf{q}_2\neq0}\sum_{\mathbf{q}_3\neq0}}
\limits_{{\mathbf q}_1+{\mathbf q}_2+{\mathbf q}_3=0}
C_3^2({\mathbf q}_1,{\mathbf q}_2, {\mathbf
q}_3)\frac{S_0(q_2)S_0(q_3)}{[1+\lambda_{q_2}S_0(q_2)][1+\lambda_{q_3}S_0(q_3)]}.
\end{eqnarray}
 
З написаного вище виразу потрібно виключити величину $\nu_0$. 
Зробити це можна з допомогою рівності \cite{VPR2007}:
\begin{eqnarray}
 c^2=\frac{N}{m}\frac{\partial^2E_0}{\partial N^2}, де
\end{eqnarray}
де $c$ --- швидкість першого звуку в бозе-системі при температурі абсолютного нуля, $N$ --- кількість частинок, $m$ --- маса частинки,
$E_0$ --- енергія основного стану в наближенні ``двох сум за хвильовим вектором''. Вираз для $E_0$ ми отримаємо трохи згодом 
і покажемо, що він збігається з відомим результатом \cite{VakUhn79_VHU79}.

Отож для величини $\nu_0$ будемо мати:
  
\begin{eqnarray}\label{nu0}
\nu_0\!&=&\!\frac{1}{\rho}\left[mc^2
\!+\!\frac{1}{8N}\!\sum_{q\neq0}\varepsilon_q\frac{1}{\alpha_q}\left(\alpha_q-\frac{1}{\alpha_q}\right)^2\!\!-\!
\frac{\hbar^2}{48mN^2}\! \mathop{\sum_{{\mathbf
q}_1\neq0}\sum_{{\mathbf q}_2\neq0}\sum_{{\mathbf q}_3\neq0}}
\limits_{{\mathbf q}_1+{\mathbf q}_2+{\mathbf
q}_3=0}\!\!\!\right.
\left\{\!2\!\left(f_1-\frac{f_2^2}{f_3}-2f_4\right)\!+\!2\!\left(f_1'-\frac{2f_2f_2'}{f_3}+\frac{f_2^2f_3'}{f_3^2}-2f_4'\right)\right.\nonumber\\
&+&\left.\left(f_1''-\frac{2f_2''+2(f_2')^2}{f_3}+
\left.\frac{4f_2f_2'f_3'+f_2^2f_3''}{f_3^2}-\frac{2f_2^2(f_3')^2}{f_3^3}-2f_4''\right)\right\}\right].\nonumber
\end{eqnarray}
 
Тут $\rho$ --- це густина бозе-системи. Явні вирази для величин $f_1,f_2,f_3,f_4,f_1',f_2',f_3',f_4',f_1'',f_2'',f_3'',f_4''$
наведені в Додатку 2.

 Якщо тепер зі знайдених середніх
$\left\langle\hat{K}_1\right\rangle$,
$\left\langle\hat{K}_{23}\right\rangle$,
$\left\langle\hat{\Phi}\right\rangle$ виділити тільки внески від парних
кореляцій і додати їх, то ми отримаємо вираз для середньої енергії в
наближенні парних кореляцій, який збігається з уже відомим \cite{VPR2007}:
\begin{eqnarray}
E&=&N\frac{mc^2}{2}+\sum_{{\mathbf
q}\neq0}\frac{\varepsilon_q}{z_0^{-1}e^{\beta\varepsilon_q}-1}+\frac{1}{2}\sum_{{\mathbf
q}\neq0}\frac{\lambda_{q}}{1+\lambda_{q}S_0(q)}\frac{\partial
S_0(q)}{\partial\beta}+\frac{1}{4}\sum_{{\mathbf
q}\neq0}\varepsilon_q(\lambda_q^2+\alpha_q-1)\frac{S_0(q)}{1+\lambda_qS_0(q)}\nonumber\\
&+&\frac{1}{2}\sum_{{\mathbf
q}\neq0}\varepsilon_q\left[\frac{\alpha_q}{\sh[\beta\alpha_q\varepsilon_q]}-\frac{1}{\sh[\beta\varepsilon_q]}\right]
+\frac{1}{16}\sum_{q\neq0}\varepsilon_q\left(1-\frac{1}{\alpha_q^2}\right)\left(\alpha_q-\frac{1}{\alpha_q}-4\alpha_q^2\right).
\end{eqnarray}
\section{Кiнетична, потенцiальна i повна енергiя в границi низьких температур}
В границі низьких температур парний структурний фактор ідеального бозе-газу рівний одиниці, а 
його похідна по оберненій температурі рівна нулю.  
Враховуючи відомі вирази для дво- і тричастинкового структурного фактора в границі низьких температур \cite{VakHryh3}, 
а також те, що
\begin{eqnarray}
\lim_{\beta\rightarrow\infty}C_2({\mathbf
q}_1)&=&\frac{1}{2}\lim_{\beta\rightarrow\infty}\widetilde{C}_2({\mathbf
q}_1)=\frac{1}{2}a_2({\mathbf
q}_1),\nonumber\\
\lim_{\beta\rightarrow\infty}C_3({\mathbf q}_1,{\mathbf q}_2,
{\mathbf
q}_3)&=&\lim_{\beta\rightarrow\infty}\widetilde{C}_3({\mathbf
q}_1,{\mathbf q}_2, {\mathbf
q}_3)=\frac{1}{6}a_3({\mathbf q}_1,{\mathbf q}_2,
{\mathbf q}_3),\nonumber\\
\lim_{\beta\rightarrow\infty}C_4({\mathbf q}_1,{\mathbf
q}_2)&=&\lim_{\beta\rightarrow\infty}\widetilde{C}_4({\mathbf
q}_1,{\mathbf q}_2)=\frac{1}{8}a_4({\mathbf q}_1,-{\mathbf q}_1,{\mathbf
q}_2,-{\mathbf q}_2),
\end{eqnarray}
де
  
\begin{eqnarray}
&&a_3{({\mathbf q}_1,{\mathbf q}_2,{\mathbf
q}_3)}=-\frac{\sum\limits_{1\leq i<j\leq3} {({\mathbf q}_i{\mathbf
q}_j)}(\alpha_{q_i}-1)(\alpha_{q_j}-1)}{2\sum\limits_{j=1}^3
{\mathbf q}_j^2\alpha_{q_j}},
\end{eqnarray}
\begin{eqnarray}
a_4{({\mathbf q}_1,-{\mathbf q}_1,{\mathbf q}_2,-{\mathbf
q}_2)}&=&\frac{{({\mathbf q}_1+{\mathbf q}_2)}^2 a_3^2{({\mathbf
q}_1+{\mathbf q}_2,-{\mathbf q}_1,-{\mathbf q}_2)}+ {({\mathbf
q}_1-{\mathbf q}_2)}^2 a_3^2{({\mathbf q}_1-{\mathbf
q}_2,-{\mathbf q}_1,{\mathbf q}_2)}}
{q_1^2\alpha_{q_1}+q_2^2\alpha_{q_2}}\nonumber\\
&-&\frac{[{({\mathbf q}_1,{\mathbf q}_2+ {\mathbf
q}_1)}(\alpha_{q_1}-1)+{({\mathbf q}_2,{\mathbf q}_1+ {\mathbf
q}_2)}(\alpha_{q_2}-1)]a_3{({\mathbf q}_1+{\mathbf q}_2,-{\mathbf
q}_1,-{\mathbf q}_2)}}
{q_1^2\alpha_{q_1}+q_2^2\alpha_{q_2}}\nonumber\\
&-&\frac{[{({\mathbf q}_1,{\mathbf q}_1- {\mathbf
q}_2)}(\alpha_{q_1}-1)+{({\mathbf q}_2,{\mathbf q}_2- {\mathbf
q}_1)}(\alpha_{q_2}-1)]a_3{({\mathbf q}_1-{\mathbf q}_2,-{\mathbf
q}_1,{\mathbf q}_2)}}
{q_1^2\alpha_{q_1}+q_2^2\alpha_{q_2}},
\end{eqnarray}
\begin{eqnarray}\label{a2}
a_2({\mathbf q}_1)&=&\frac{1}{N}\sum_{{\mathbf q}_2\neq0}
\left[\frac{q_2^2}{2q_1^2\alpha_{q_1}} a_4{({\mathbf
q}_1,-{\mathbf q}_1,{\mathbf q}_2,-{\mathbf q}_2)}+
\frac{({\mathbf q}_2,{\mathbf q}_1+{\mathbf q}_2)}
{q_1^2\alpha_{q_1}}a_3{({\mathbf q}_1,{\mathbf q}_2,-{\mathbf
q}_1-{\mathbf q}_2)} \right],
\end{eqnarray}
 
для середньої кінетичної 
енергії в границі низьких температур знайдемо такий вираз:
  
\begin{eqnarray}
\left\langle\hat{K}\right\rangle&=&\frac{1}{4}\sum_{{\mathbf
q}\neq0}\frac{\hbar^2q^2}{2m}\frac{(\alpha_{q}-1)^2}{\alpha_{q_1}}
+\frac{1}{8N}\mathop{\sum_{\mathbf{q}_1\neq0}\sum_{\mathbf{q}_2\neq0}\sum_{\mathbf{q}_3\neq0}}
\limits_{{\mathbf q}_1+{\mathbf q}_2+{\mathbf
q}_3=0}\frac{\hbar^2q_1^2}{2m}\frac{\alpha_{q_1}-1}{\alpha_{q_1}}\left[\prod_{i=1}^3\frac{\alpha_{q_i}-1}{\alpha_{q_i}}\right]-
\frac{1}{2N}\sum_{{\mathbf
q_1}\neq0}\frac{\hbar^2q_1^2}{2m}\frac{\alpha_{q_1}^2-1}{\alpha_{q_1}^2}a_2({\mathbf
q}_1)\nonumber\\
&-&\frac{1}{2N}\mathop{\sum_{\mathbf{q}_1\neq0}\sum_{\mathbf{q}_2\neq0}\sum_{\mathbf{q}_3\neq0}}
\limits_{{\mathbf q}_1+{\mathbf q}_2+{\mathbf
q}_3=0}\frac{\hbar^2q_1^2}{2m}\frac{\alpha_{q_1}-1}{\alpha_{q_1}^2\alpha_{q_2}\alpha_{q_3}}a_3{({\mathbf
q}_1,{\mathbf q}_2,{\mathbf
q}_3)}-\frac{1}{4N}\sum_{\mathbf{q}_1\neq0}\sum_{\mathbf{q}_2\neq0}\frac{\hbar^2q_1^2}{2m}
\frac{\alpha_{q_1}^2-1}{\alpha_{q_1}^2\alpha_{q_2}}a_4{({\mathbf
q}_1,-{\mathbf q}_1,{\mathbf q}_2,-{\mathbf q}_2)}\nonumber\\
&+&\frac{1}{2N}\mathop{\sum_{\mathbf{q}_1\neq0}\sum_{\mathbf{q}_2\neq0}\sum_{\mathbf{q}_3\neq0}}
\limits_{{\mathbf q}_1+{\mathbf q}_2+{\mathbf
q}_3=0}\frac{\hbar^2q_1^2}{2m}\frac{1}{\alpha_{q_1}^2\alpha_{q_2}\alpha_{q_3}}a_3^2{({\mathbf
q}_1,{\mathbf q}_2,{\mathbf
q}_3)}-\frac{1}{4N}\sum_{\mathbf{q}_1\neq0}\sum_{\mathbf{q}_2\neq0}\frac{\hbar^2(\mathbf{q}_1\mathbf{q}_2)}{2m}
\left(\frac{(\alpha_{q_1}-1)(\alpha_{q_2}-1)}{\alpha_{q_1}\alpha_{q_2}\alpha_{|\mathbf{q}_1+\mathbf{q}_2|}}\right.\nonumber\\
&-&\left.2\frac{\alpha_{q_1}\alpha_{q_2}-1}{\alpha_{q_1}\alpha_{q_2}\alpha_{q_3}}a_3{({\mathbf q}_1,{\mathbf q}_2,{\mathbf
q}_3)}\right)
\end{eqnarray}
Аналогічно, для середньої потенціальної енергії будемо мати:
\begin{eqnarray}
\left\langle\hat{\Phi}\right\rangle&=&N\frac{mc^2}{2}+
\frac{1}{16}\sum_{q\neq0}\frac{\hbar^2q^2}{2m}\left(1-\frac{1}{\alpha_q^2}\right)\left(\alpha_q-\frac{1}{\alpha_q}-4\alpha_q^2\right)+\sum_{{\mathbf
q}\neq0}\frac{\hbar^2q^2}{8m}\frac{\alpha_{q}^2-1}{\alpha_{q_1}}\nonumber\\
&+&\frac{1}{8N}\mathop{\sum_{\mathbf{q}_1\neq0}\sum_{\mathbf{q}_2\neq0}\sum_{\mathbf{q}_3\neq0}}
\limits_{{\mathbf q}_1+{\mathbf q}_2+{\mathbf
q}_3=0}\frac{\hbar^2q_1^2}{2m}\frac{\alpha_{q_1}+1}{\alpha_{q_1}}
\left[\prod_{i=1}^3\frac{\alpha_{q_i}-1}{\alpha_{q_i}}\right]+\frac{1}{2N}\sum_{{\mathbf
q_1}\neq0}\frac{\hbar^2q_1^2}{2m}\frac{\alpha_{q_1}^2-1}{\alpha_{q_1}^2}a_2({\mathbf
q}_1)\nonumber\\
&+&\frac{1}{2N}\mathop{\sum_{\mathbf{q}_1\neq0}\sum_{\mathbf{q}_2\neq0}\sum_{\mathbf{q}_3\neq0}}
\limits_{{\mathbf q}_1+{\mathbf q}_2+{\mathbf
q}_3=0}\frac{\hbar^2q_1^2}{2m}\frac{\alpha_{q_1}^2-1}{\alpha_{q_1}^2\alpha_{q_2}\alpha_{q_3}}a_3{({\mathbf
q}_1,{\mathbf q}_2,{\mathbf
q}_3)}+\frac{1}{4N}\sum_{\mathbf{q}_1\neq0}\sum_{\mathbf{q}_2\neq0}\frac{\hbar^2q_1^2}{2m}
\frac{\alpha_{q_1}^2-1}{\alpha_{q_1}^2\alpha_{q_2}}a_4{({\mathbf
q}_1,-{\mathbf q}_1,{\mathbf q}_2,-{\mathbf q}_2)}\nonumber\\
&+&\frac{1}{2N}\mathop{\sum_{\mathbf{q}_1\neq0}\sum_{\mathbf{q}_2\neq0}\sum_{\mathbf{q}_3\neq0}}
\limits_{{\mathbf q}_1+{\mathbf q}_2+{\mathbf
q}_3=0}\frac{\hbar^2q_1^2}{2m}\frac{\alpha_{q_1}^2-1}{\alpha_{q_1}^2\alpha_{q_2}\alpha_{q_3}}a_3^2{({\mathbf
q}_1,{\mathbf q}_2,{\mathbf q}_3)}.
\end{eqnarray}
Додавши вирази для середніх значень кінетичної і потенціальної енергій та провівши необхідні перетворення,
ми отримаємо величину середнього
значення повної енергії в границі низьких температур в такому вигляді:
\begin{eqnarray}
E&=&\left\langle\hat{K}\right\rangle+\left\langle\hat{\Phi}\right\rangle=N\frac{mc^2}{2}
-\frac{1}{16}\sum_{q\neq0}\frac{\hbar^2q^2}{2m}\left(1-\frac{1}{\alpha_q^2}\right)^3\left(\alpha_q^2-\frac{3}{4}\alpha_q+\frac{1}{4}\right)
+\frac{\hbar^2}{48mN} \mathop{\sum_{{\mathbf
q}_1\neq0}\sum_{{\mathbf q}_2\neq0}\sum_{{\mathbf q}_3\neq0}}
\limits_{{\mathbf q}_1+{\mathbf q}_2+{\mathbf q}_3=0}
\left[(q_1^2+q_2^2+q_3^2)\right.\nonumber\\
&\times&\left.\left(1-\frac{1}{\alpha_{q_1}}\right)\left(1-\frac{1}{\alpha_{q_2}}\right)\left(1-\frac{1}{\alpha_{q_3}}\right)
-\frac{\left(\sum\limits_{1\leq i<j\leq3}({\mathbf q}_i{\mathbf
q}_j)(\alpha_{q_i}-1)(\alpha_{q_j}-1)\right)^2}
{\alpha_{q_1}\alpha_{q_2}\alpha_{q_3}\sum\limits_{j=1}^3q_j^2\alpha_{q_j}}\right]\nonumber\\
&-&\frac{\hbar^2}{24mN}\mathop{\sum_{{\mathbf q}_1\neq0}
\sum_{{\mathbf q}_2\neq0}\sum_{{\mathbf q}_3\neq0}}
\limits_{{\mathbf q}_1+{\mathbf q}_2+{\mathbf q}_3=0}\sum_{1\leq
i<j\leq3}{({\mathbf q}_i{\mathbf
q}_j)}\left(1-\frac{1}{\alpha_{q_i}}\right)
\left(1-\frac{1}{\alpha_{q_j}}\right).
\end{eqnarray}
 
 Цей вираз для основного стану багатобозонної системи збігається зі знайденим раніше \cite{VakUhn79_VHU79}. 

\section{Кiнетична, потенцiальна i повна енергiя в границi високих температур}

В границі високих температур парний, три- і чотиричастинковий
структурні фактори багатобозонної системи переходять у відповідні
вирази для ідеального бозе-газу, а всі внески від три- і чотиричастинкових кореляцій стають 
рівними нулю \cite{VakHryh3}. Тому пошук значення кінетичної енергії в границі високих температур зводиться до наступної задачі:
\begin{eqnarray}\label{Vz0}
\lim_{\beta\rightarrow0}\left\langle\hat{K}\right\rangle=\lim_{\beta\rightarrow0}\sum_{{\mathbf
q}\neq0}\frac{\varepsilon_q}{z_0^{-1}e^{\beta\varepsilon_q}-1},
\end{eqnarray}
оскільки всі інші доданки у згаданій вище границі рівні нулю.
Вираз (\ref{Vz0}) містить величину $z_0=\exp[\beta\mu]$
($\mu$ --- це хімічний потенціал),
 яка називається активність. Умову для її визначення можна записати цим рівнянням:
 \begin{eqnarray}
 \sum_{{\bf q}\neq0}\frac{1}{z_0^{-1}e^{\beta\varepsilon_q}-1}=N,
\end{eqnarray}
де $N$ --- число частинок в бозе-системі. В границі високих
температур $z_0\rightarrow0$, тому написану вище умову можна
подати в такий спосіб:
\begin{eqnarray}
 z_0\sum_{{\bf q}\neq0}e^{-\beta\varepsilon_q}=N.
\end{eqnarray}
Переходячи в цій рівності від пісумовування до інтегрування,
будемо мати:
\begin{eqnarray}
z_0\frac{4\pi V}{(2\pi)^3}\int\limits_0^\infty
q^2e^{-\beta\frac{\hbar^2}{2m}q^2}dq=
z_0\frac{V}{4}\left(\frac{2m}{\pi\hbar^2}\right)^{\frac{3}{2}}\!\!T^{\frac{3}{2}}=N.
\end{eqnarray}
Звідси
\begin{eqnarray}
z_0^{-1}=\frac{1}{4\rho}\left(\frac{2m}{\pi\hbar^2}\right)^{\frac{3}{2}}T^{\frac{3}{2}}.
\end{eqnarray}
Підставивши щойно отриманий результат у рівність (\ref{Vz0}), провівши відповідні спрощення і перейшовши 
від підсумовування до інтегрування, знайдемо, що
\begin{eqnarray}
\lim_{\beta\rightarrow0}\left\langle\hat{K}\right\rangle=\frac{3}{2}NT.
\end{eqnarray}
Ми отримали результат, котрий збігається з виразом для ідеального
газу.

Для значення середньої потенціальної енергії в границі високих
температур будемо мати:
\begin{eqnarray}
\lim_{\beta\rightarrow0}\left\langle\Phi\right\rangle&=&\frac{N(N-1)}{2V}\nu_0+\lim_{\beta\rightarrow0}\frac{N}{2V}\sum_{{\mathbf
q}\neq0}\nu_q (\left\langle\rho_{{\mathbf q}}\rho_{-{\mathbf q}}\right\rangle-1)=\frac{N(N-1)}{2V}\nu_0,
\end{eqnarray}
оскільки
\begin{eqnarray}
\lim_{\beta\rightarrow0}\left\langle\rho_{{\mathbf
q}}\rho_{-{\mathbf q}}\right\rangle=1.
\end{eqnarray}
Об'єднуючи отримані результати, для внутрішньої енергії багатобозонної системи в 
границі високих температур здобудемо такий вираз:
\begin{eqnarray}
\lim_{\beta\rightarrow0}E&=&\lim_{\beta\rightarrow0}\left\langle\hat{K}\right\rangle+
\lim_{\beta\rightarrow0}\left\langle\Phi\right\rangle=\frac{3}{2}NT+\frac{N(N-1)}{2V}\nu_0,
\end{eqnarray}
який  збігається з результатом роботи \cite{Vak2004}.

\section{Чисельні розрахунки}
Для чисельного розрахунку внутрішньої енергії рідкого $^4$He скористаємося знайденими  виразами для $\langle K_1\rangle$
(\ref{K1_res}),  $\langle K_{23}\rangle$ (\ref{K23_res}) і  $\langle\Phi\rangle$ (\ref{Penergy_res}). Величину $\nu_0$ будемо 
шукати за наведеною вище формулою (\ref{nu0}). Щоб виконати поставлену задачу 
потрібно від підсумовування перейти до інтегрування за відомим правилом \cite{Vstup}:
\begin{eqnarray}
\sum_{{\bf k}}=\frac{V}{(2\pi)^3}\int d{\bf k}.
\end{eqnarray}
У нашому випадку двох сум це правило стане дещо складнішим:
\begin{eqnarray}
&&\frac{1}{N^2}\mathop{\sum_{\mathbf{k}_1\neq0}\sum_{\mathbf{k}_2\neq0}\sum_{\mathbf{k}_3\neq0}}
\limits_{{\mathbf k}_1+{\mathbf k}_2+{\mathbf k}_3=0}=\frac{1}{8\pi^4\rho^2}
\int\limits_0^\infty k_1dk_1\int\limits_0^\infty k_2dk_2\int\limits_{|q_1-k_2|}^{|q_1+k_2|}k_3dk_3,
\end{eqnarray}
де $\rho$ --- це рівноважна густина бозе-системи. Для такої
квантової рідини як $^4He$ вона рівна $\rho=0.02185$\AA$^{-3}$; 
швидкість першого звуку в рідкому $^4He$ при нулі температур  $c=238.2$ м/с
\cite{DonBar}.

При чисельних розрахунках кінетичної, потенціальної і повної внутрішньої енергії у виразах, які відтворюють 
наближення парних кореляцій, ми використовували ефективну масу атома $^4$He в рідині \cite{HP2014}, 
натомість у виразах, які містять дві суми за хвильовим вектором, стоїть ``гола'' маса. 
Обгрунтування такого підходу наведено в  роботі \cite{VakHryh3}.

\newpage 
\begin{figure}[h!]
\center{\includegraphics[scale=0.85]{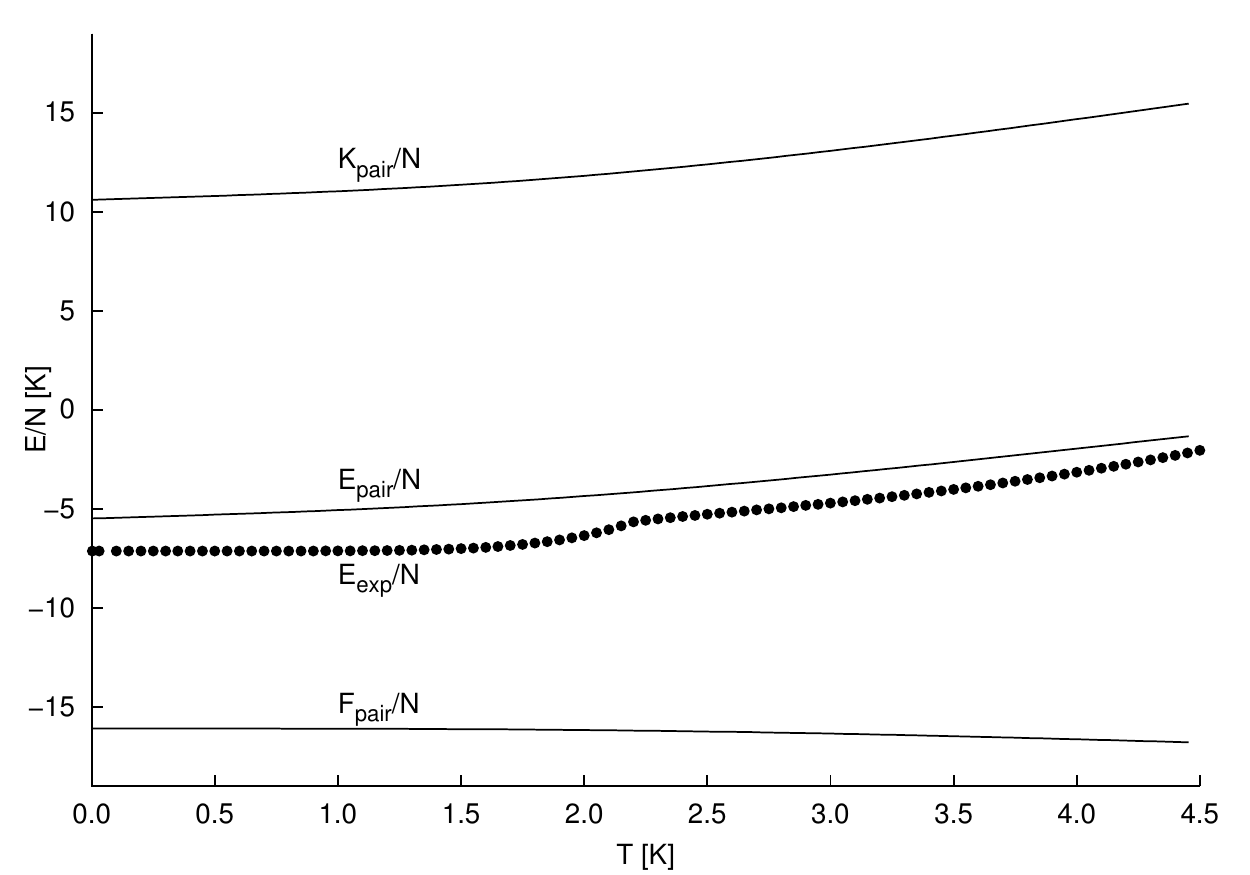}}
\caption{Температурна залежність кінетичної, потенціальної і повної внутрішньої енергії рідкого $^4$He, 
розрахованих в наближенні парних кореляцій. $K_{pair}/N$ --- кінетична енергія, $F_{pair}/N$ --- потенціальна енергія, 
$E_{pair}/N$ --- повна внутрішня 
енергія,
$E_{exp}/N$ --- експериментальні дані.}
\end{figure}

\begin{figure}[h!]
\center{\includegraphics[scale=0.9]{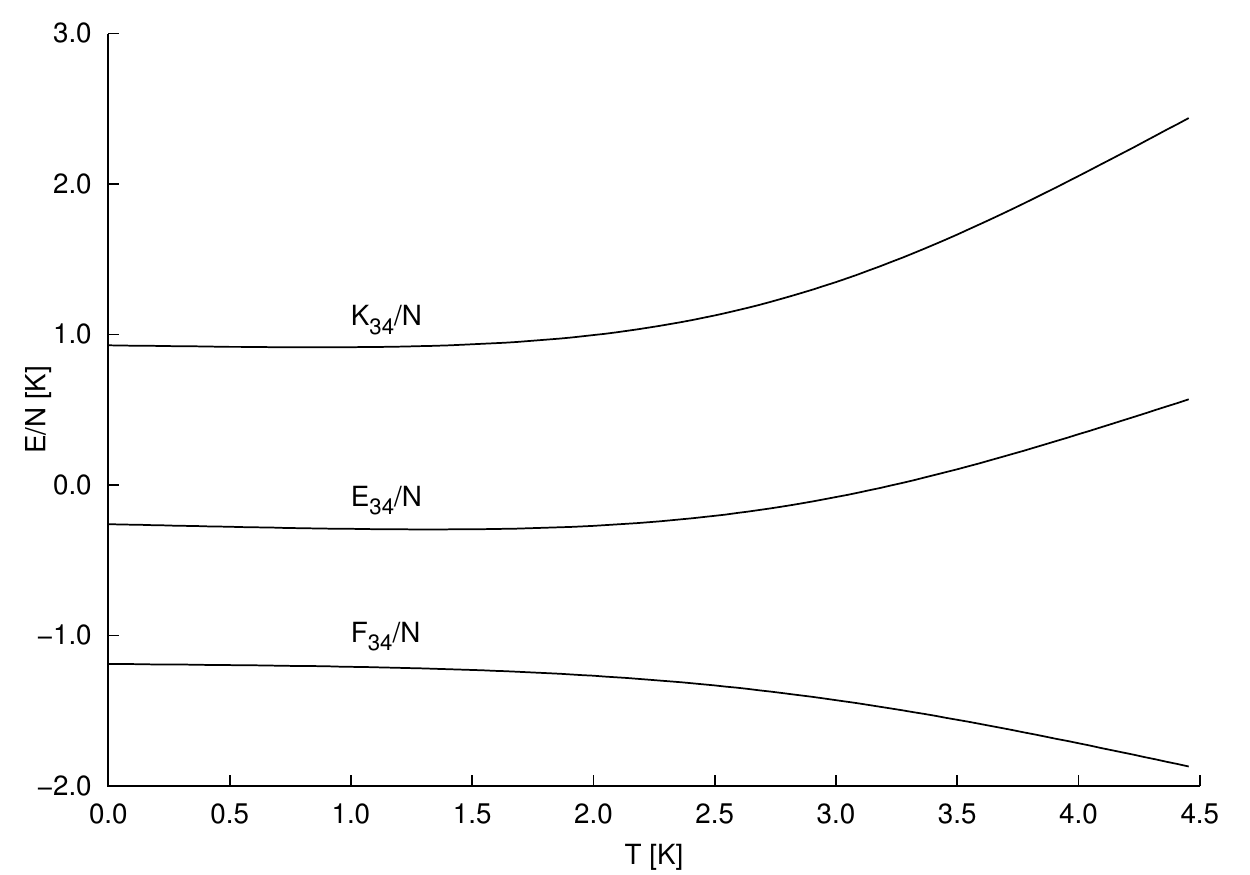}}
\caption{Внесок три- ча чотиричастинкових кореляцій в зачення кінетичної, потенціальної і повної внутрішньої енергії
рідкого $^4$He.
$K_{34}/N$ --- внесок у кінетичну енергію, $F_{34}/N$ --- внесок у потенціальну енергію, $E_{34}/N$ --- внесок у повну внутрішню енергію.} 
\end{figure}

\begin{figure}[h!]
\center{\includegraphics[scale=1.1]{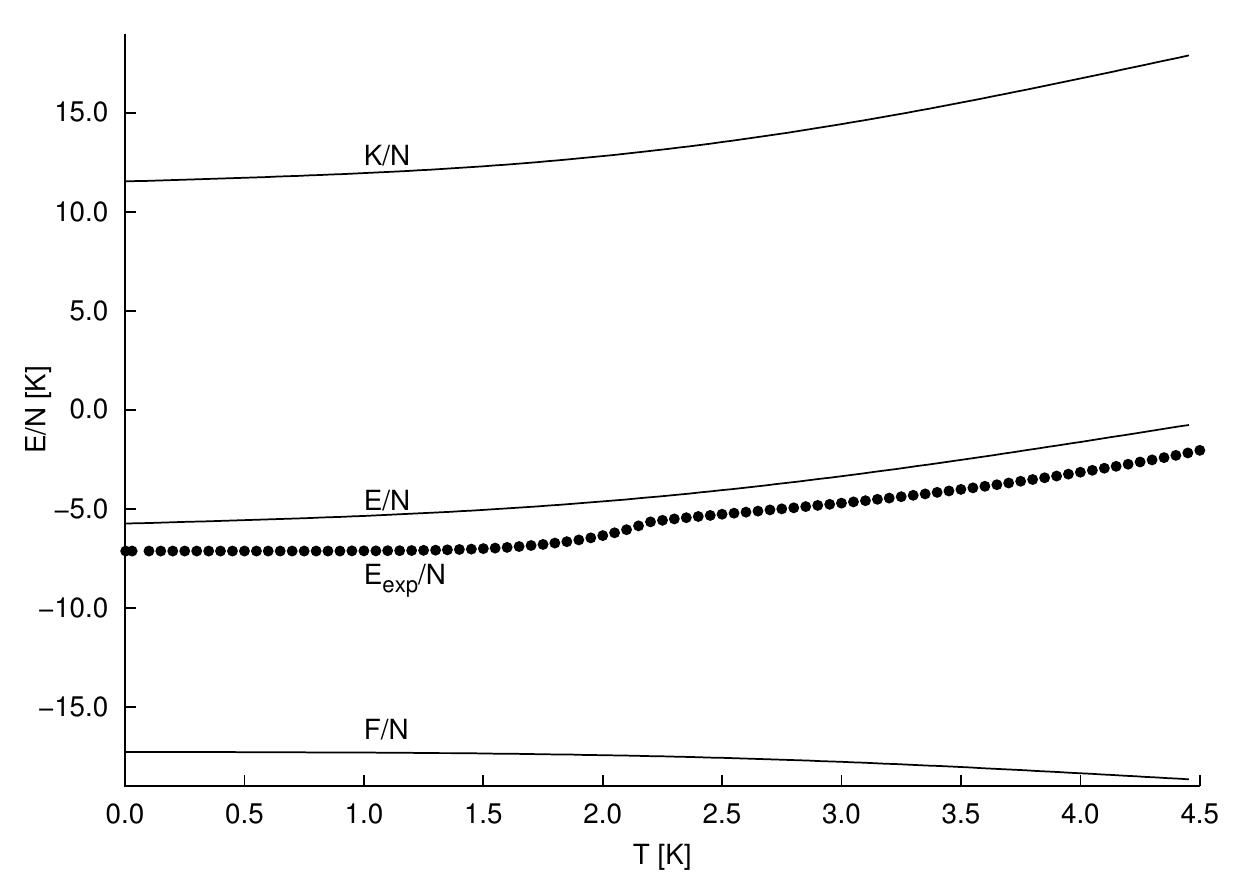}}
\caption{Температурна залежність кінетичної, потенціальної і повної внутрішньої енергії рідкого $^4$He, 
розрахованих в наближенні ``двох сум за хвильовим вектором''.  $K/N$ --- кінетична енергія, 
$F/N$ --- потенціальна енергія, $E/N$ --- повна внутрішня енергія,
$E_{exp}/N$ --- експериментальні дані.}
\end{figure}

\begin{figure}[h!]
\center{\includegraphics[scale=1.0]{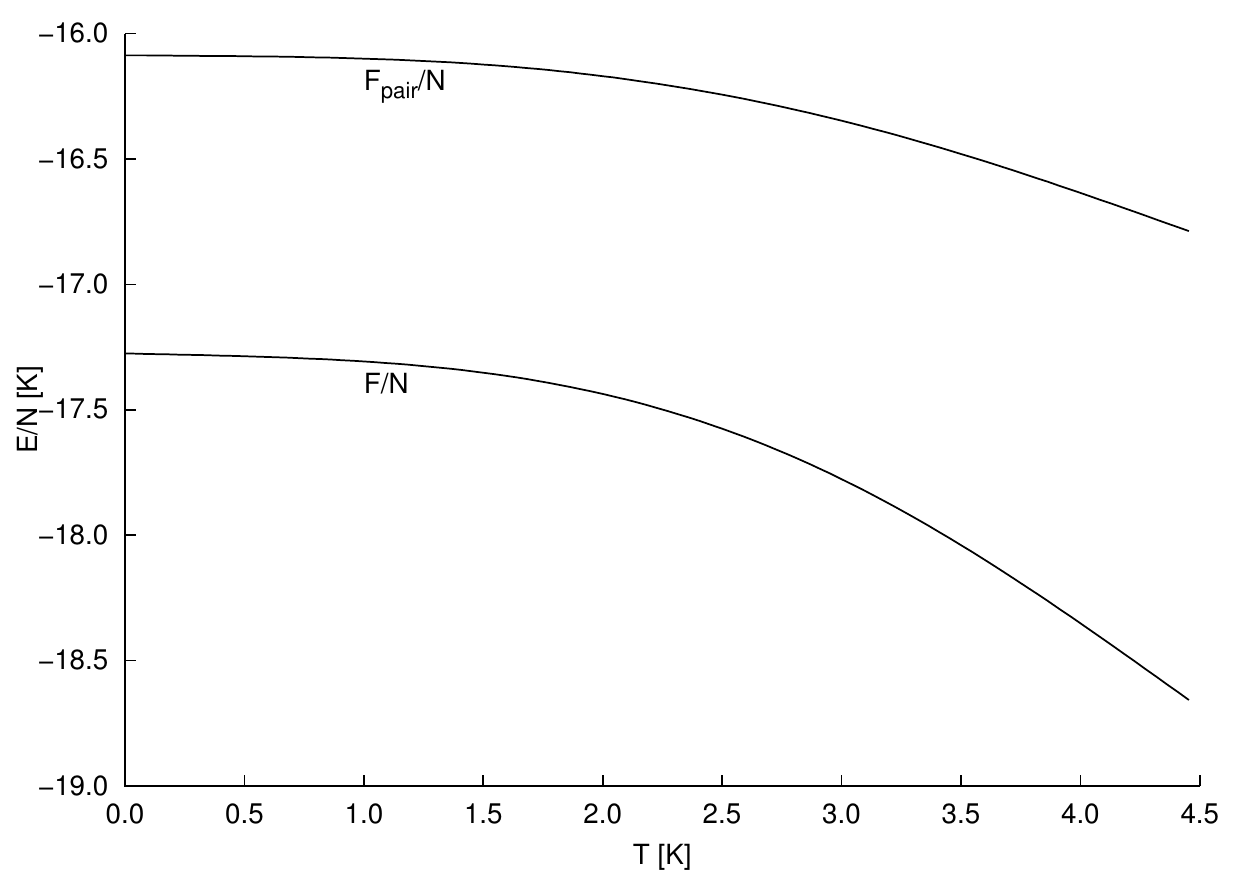}}
\caption{Температурна залежність потенціальної енергії рідкого $^4$He. $F_{pair}/N$ --- потенціальна енергія в наближенні парних кореляцій,
$F/N$ --- потенціальна енергія в наближенні ``двох сум за хвильовим вектором''.}
\end{figure}

\begin{figure}[h!]
\includegraphics[scale=1.0]{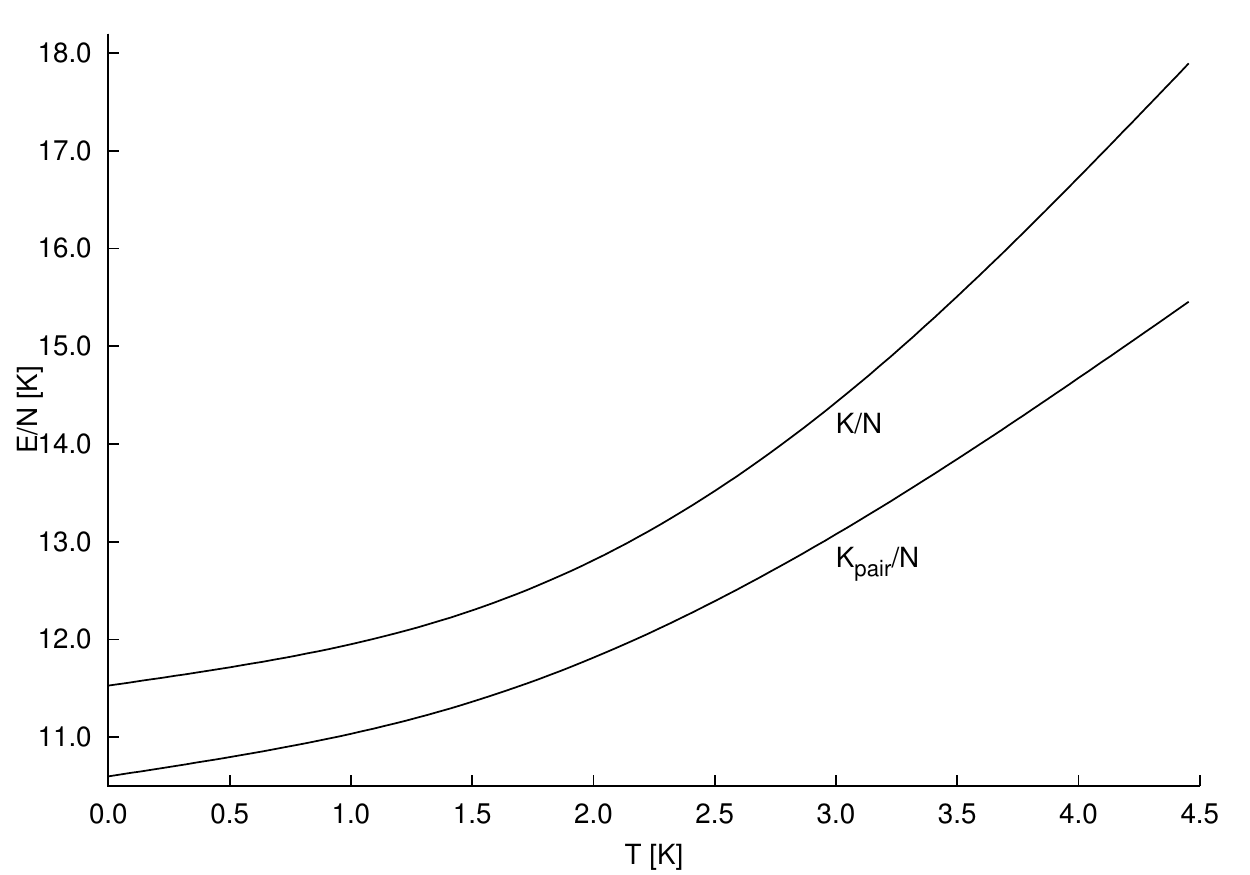}
\caption{Температурна залежність кінетичної енергії рідкого $^4$He. $K_{pair}/N$ --- кінетична енергія в наближенні парних кореляцій,
$K/N$ --- кінетична енергія в наближенні ``двох сум за хвильовим вектором''.}
\end{figure}

\begin{figure}[h!]
\center{\includegraphics[scale=1.0]{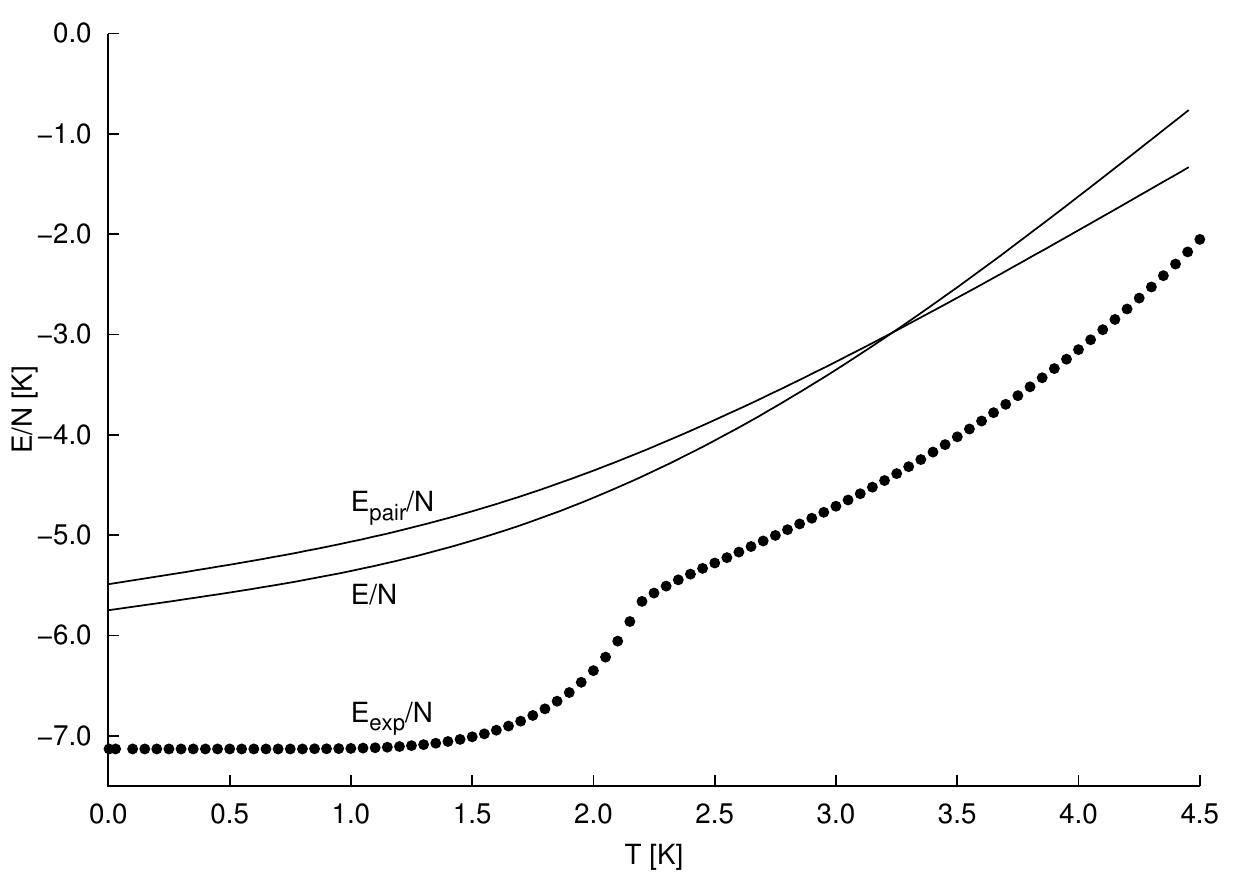}}
\caption{Температурна залежність  повної внутрішньої енергії рідкого $^4He$.
$E_{pair}/N$ --- внутрішня енергія в наближенні парних кореляцій,
$E/N$ --- внутрішня енергія в наближенні ``двох сум за хвильовим вектором".}
\end{figure}

\newpage
\section{Висновок}
 В цій роботі були знайдені вирази для кінетичної,
потенціальної та повної енергії в наближенні ``двох сум за
хвильовим вектором'' із врахуванням прямих три- і чотиричастинкових
кореляцій в широкотемпературній ділянці. В границі низьких
температур  для внутрішньої енергії ми отримали вже відоме значення
\cite{VakUhn79_VHU79}. Ці
самі слова будуть актуальними і по відношенню до границі високих
температур.   Отримані вирази є доволі громіздкими. Для їх аналізу
були застосовані чисельні методи і в результаті здобуто
графічне представлення температурної залежності кінетичної, потенціальної і повної енергії
рідкого $^4$He.
Результати, які ми отримали в наближенні ``двох сум за хвильовим вектором'' 
краще узгоджуються з експериментальними даними в порівнянні з наближенням парних кореляцій.
   
\section{Додаток 1}
\begin{eqnarray}
&&\widetilde{C}_2({\bf q_1})=2C_2({\bf q_1})+\frac{1}{2}\sum_{{\bf
q_2}\neq0}\frac{\hbar^2}{2m}\frac{q_1^2+q_2^2}{\alpha_{q_2}\sh^2\left[\beta
E_{q_1}\right]\sh[\beta E_2]}\left\{\frac{\beta}{2}\ch[\beta
E_{q_1}]\ch[\beta
E_{q_2}]\right.\nonumber\\
&-&\left.\frac{\ch[\beta E_1]\sh[\beta E_2]}{2E_2}-\frac{\sh[\beta
E_1]\ch[\beta
E_2]}{2E_1}+\frac{\sh[\beta(E_1+E_2)]}{4(E_1+E_2)}+\frac{\sh[\beta(E_1-E_2)]}{4(E_1-E_2)}\right\}
\nonumber\\
&-&\frac{1}{8}\mathop{\sum_{\mathbf{q}_2\neq0}\sum_{\mathbf{q}_3\neq0}}\limits_{\mathbf{q}_1+\mathbf{q}_2+\mathbf{q}_3=0}
\left(\frac{\hbar^2}{2m}\right)^2\frac{Q(\tilde{\alpha}_{q_1},\tilde{\alpha}_{q_2},
\alpha_{q_3})} {\alpha_{q_2}\alpha_{q_3}\tilde{E}\sh^2\left[\beta
E_{q_1}\right]\sh\left[\beta E_{q_2}\right]\sh\left[\beta
E_{q_3}\right]}\nonumber\\
&\times& \left\{\frac{\beta}{4}\ch\left[\beta
\tilde{E}_{q_1}\right]\sh\left[\beta(\tilde{E}_{q_2}+E_{q_3})\right]Q(\tilde{\alpha}_{q_1},\tilde{\alpha}_{q_2},
\alpha_{q_3})-\frac{\sh\left[\frac{\beta}{2}\tilde{E}\right]}{2\tilde{E}}\right.\nonumber\\
&\times&\left(\sh\left[\frac{\beta}{2}\tilde{E}\right]+
\sh\left[\frac{\beta}{2}(-\tilde{E}_{q_1}+\tilde{E}_{q_2}+E_{q_3})\right]\ch\left[\beta
\tilde{E}_{q_1}\right]\right)Q(\tilde{\alpha}_{q_1},\tilde{\alpha}_{q_2},
\alpha_{q_3})\nonumber\\
&+&\left[-\frac{\sh\left[\frac{\beta}{2}\tilde{E}_{q_2}\right]}{2\tilde{E}_{q_2}}
\left(\ch\left[\beta\tilde{E}_{q_1}\right]\sh\left[\frac{\beta}{2}(\tilde{E}_{q_2}+2E_{q_3})\right]+
\sh\left[\frac{\beta}{2}\tilde{E}_{q_2}\right]\right)+
\frac{\sh\left[\frac{\beta}{2}(\tilde{E}_{q_1}+E_{q_3})\right]}{2(\tilde{E}_{q_1}+E_{q_3})}\right.\nonumber\\
&\times&\left.
\left(\ch\left[\beta\tilde{E}_{q_1}\right]\sh\left[\frac{\beta}{2}(E_{q_3}-\tilde{E}_{q_1})\right]+
\sh\left[\frac{\beta}{2}(\tilde{E}_{q_1}+2\tilde{E}_{q_2}+E_{q_3})\right]\right)\right]Q(\tilde{\alpha}_{q_1},-\tilde{\alpha}_{q_2},
\alpha_{q_3})\nonumber\\
&+&\left[-\frac{\sh\left[\frac{\beta}{2}\tilde{E}_{q_1}\right]}{2\tilde{E}_{q_1}}
\left(-\ch\left[\beta\tilde{E}_{q_1}\right]\sh\left[\frac{\beta}{2}\tilde{E}_{q_1}\right]+
\sh\left[\frac{\beta}{2}(\tilde{E}_{q_1}+2\tilde{E}_{q_2}+2E_{q_3})\right]\right)\right.\nonumber\\
&+&\left.
\frac{\sh\left[\frac{\beta}{2}(\tilde{E}_{q_2}+E_{q_3})\right]}{2(\tilde{E}_{q_2}+E_{q_3})}
\ch^2\left[\frac{\beta}{2}\tilde{E}_{q_1}\right]\sh\left[\frac{\beta}{2}(\tilde{E}_{q_2}+E_{q_3})\right]\right]Q(-\tilde{\alpha}_{q_1},\tilde{\alpha}_{q_2},
\alpha_{q_3})\nonumber\\
&+&\left[-\frac{\sh\left[\frac{\beta}{2}E_{q_3}\right]}{2E_{q_3}}
\left(\ch\left[\beta\tilde{E}_{q_1}\right]\sh\left[\frac{\beta}{2}(2\tilde{E}_{q_2}+E_{q_3})\right]+\sh\left[\frac{\beta}{2}E_{q_3}\right]\right)+
\frac{\sh\left[\frac{\beta}{2}(\tilde{E}_{q_1}+\tilde{E}_{q_2})\right]}{2(\tilde{E}_{q_1}+\tilde{E}_{q_2})}\right.\nonumber\\
&\times&\left.\left.
\left(\ch\left[\beta\tilde{E}_{q_1}\right]\sh\left[\frac{\beta}{2}(\tilde{E}_{q_2}-\tilde{E}_{q_1})\right]+
\sh\left[\frac{\beta}{2}(\tilde{E}_{q_1}+\tilde{E}_{q_2}+2E_{q_3})\right]\right)\right]Q(-\tilde{\alpha}_{q_1},-\tilde{\alpha}_{q_2},
\alpha_{q_3})\right\}.\nonumber\\
\end{eqnarray}
\begin{eqnarray}
&&\widetilde{C}_3({\bf q_1},{\bf q_2},{\bf
q_3})=-\frac{1}{4}\frac{\hbar^2}{2m}\frac{(\bf{q}_1\bf{q}_2)}{\sh[\beta
E_{q_1}]\sh[\beta E_{q_2}]\sh[\beta
E_{q_3}]}\sum_{\pm_1}\sum_{\pm_2}\left(\alpha_{q_1}\alpha_{q_2}\pm_1\pm_21\right)\nonumber\\
&\times&
\sh\left[\frac{\beta}{2}\left(\tilde{E}_{q_1}+\tilde{E}_{q_2}\right)\right]\ch\left[\frac{\beta}{2}E_{q_3}\right]
\frac{\sh\left[\frac{\beta}{2}\left(\tilde{E}_{q_1}+
\tilde{E}_{q_2}+E_{q_3}\right)\right]}{\tilde{E}_{q_1}+\tilde{E}_{q_2}+E_{q_3}}.
\end{eqnarray}
\begin{eqnarray}
&&\widetilde{C}_4({\bf q_1},{\bf q_2})=C_4({\bf q_1},{\bf q_2})-\frac{1}{64}\sum_{{\bf
q_1}\neq0}\sum_{{\bf
q_2}\neq0}\frac{\hbar^2}{2m}\frac{(q_1^2+q_2^2)}{\ch^2\left[\frac{\beta}{2}
E_1\right]\sh^2\left[\beta
E_2\right]}\left\{2\beta\ch\left[\beta
E_{q_2}\right]+\frac{2\sh[\beta
E_2\!]}{E_2}\right.\nonumber\\
&-&\frac{2\sh[\beta E_1]\ch[\beta
E_2]}{E_1}\!+\left.\frac{\sh[\beta(E_1\!+\!E_2)]}{(E_1\!+\!E_2)}\!+\!\frac{\sh[\beta(E_1-E_2)]}{(E_1-E_2)}\right\}-
\frac{1}{64}\!\!\!\!\mathop{\sum_{\mathbf{q}_3\neq0}}\limits_{\mathbf{q}_1+\mathbf{q}_2+\mathbf{q}_3=0}
\!\!\!\!
\left(\frac{\hbar^2}{2m}\right)^2\nonumber\\
&\times&\!\frac{Q(\tilde{\alpha}_{q_1},\tilde{\alpha}_{q_2},
\alpha_{q_3})} {\alpha_{q_3}\tilde{E}\sh^2\left[\beta
E_1\right]\sh^2\left[\beta E_2\right]\sh\left[\beta
E_3\right]}\left\{\left(\frac{\beta}{2}\ch\left[\beta
\tilde{E}_{q_2}\right]\sh\left[\beta
E_{q_3}\right]-\frac{\sh\left[\frac{\beta}{2}\tilde{E}\right]}{2\tilde{E}}\right.\right.\nonumber\\
&\times&\left(\ch\left[\frac{\beta}{2}\tilde{E}_{q_2}\right]\sh\left[\frac{\beta}{2}(\tilde{E}_{q_1}+\tilde{E}_{q_2}-E_{q_3})\right]+
\ch\left[\frac{\beta}{2}\tilde{E}\right]\right)Q(\tilde{\alpha}_{q_1},\tilde{\alpha}_{q_2},
\alpha_{q_3})\nonumber\\
&+&\left[\frac{\sh\left[\frac{\beta}{2}\tilde{E}_{q_2}\right]}{\tilde{E}_{q_2}}
\left(\ch\left[\beta\tilde{E}_{q_2}\right]\sh\left[\frac{\beta}{2}\tilde{E}_{q_2}\right]+
\sh\left[\frac{\beta}{2}(\tilde{E}_{q_2}+2E_{q_3})\right]\right)-
\frac{\sh\left[\frac{\beta}{2}(\tilde{E}_{q_1}+E_{q_3})\right]}{(\tilde{E}_{q_1}+E_{q_3})}\right.\nonumber\\
&\times&\left.
\left(\ch\left[\beta\tilde{E}_{q_2}\right]\sh\left[\frac{\beta}{2}(\tilde{E}_{q_1}+E_{q_3})\right]+
\sh\left[\frac{\beta}{2}(\tilde{E}_{q_1}-E_{q_3})\right]\right)\right]Q(\tilde{\alpha}_{q_1},-\tilde{\alpha}_{q_2},
\alpha_{q_3})\nonumber\\
&+&\left[\frac{\sh\left[\frac{\beta}{2}\tilde{E}_{q_1}\right]}{\tilde{E}_{q_1}}
\left(\ch\left[\beta\tilde{E}_{q_2}\right]\sh\left[\frac{\beta}{2}\tilde{E}_{q_1}+2E_{q_3}\right]+
\sh\left[\frac{\beta}{2}\tilde{E}_{q_1}\right]\right)-
\frac{\sh\left[\frac{\beta}{2}(\tilde{E}_{q_2}+E_{q_3})\right]}{(\tilde{E}_{q_2}+E_{q_3})}\right.\nonumber\\
&\times&\left.
\sh\left[\beta\tilde{E}_{q_2}\right]\ch\left[\frac{\beta}{2}(\tilde{E}_{q_2}-E_{q_3})\right]\right]Q(-\tilde{\alpha}_{q_1},\tilde{\alpha}_{q_2},
\alpha_{q_3})
+\left[\frac{\sh^2\left[\frac{\beta}{2}E_{q_3}\right]}{E_{q_3}}
\sh^2\left[\frac{\beta}{2}\tilde{E}_{q_2}\right]-
\frac{\sh\left[\frac{\beta}{2}(\tilde{E}_{q_1}+\tilde{E}_{q_2})\right]}{(\tilde{E}_{q_1}+\tilde{E}_{q_2})}\right.\nonumber\\
&\times&\left.
\left(\ch\left[\beta\tilde{E}_{q_2}\right]\sh\left[\frac{\beta}{2}(\tilde{E}_{q_1}+\tilde{E}_{q_2})\right]+
\sh\left[\frac{\beta}{2}(\tilde{E}_{q_1}+\tilde{E}_{q_1}+2E_{q_3})\right]\right)\right]
Q(\tilde{\alpha}_{q_1},-\tilde{\alpha}_{q_2},
\left.\alpha_{q_3})\frac{}{}\right\}.\nonumber\\
\end{eqnarray}
В написаних вище виразах введені такі позначення:
\begin{eqnarray}
\tilde{E}_{q_1}=\pm_1E_{q_1};\quad
\tilde{E}_{q_2}=\pm_2E_{q_1};\quad
\tilde{\alpha}_{q_1}=\pm_1\alpha_{q_1};\quad
\tilde{\alpha}_{q_2}=\pm_1\alpha_{q_2};\quad
\tilde{E}=\pm_1E_{q_1}\pm_2E_{q_2}+E_{q_3};\nonumber
\end{eqnarray}
\begin{eqnarray}
Q(\tilde{\alpha}_{q_1},\tilde{\alpha}_{q_2},\alpha_{q_3})=(\pm_1\pm_2\alpha_{q_1}\alpha_{q_2}+1)({\bf
q}_1{\bf q}_2)+(\pm_1\alpha_{q_1}\alpha_{q_3}+1)({\bf q}_1{\bf
q}_3)+(\pm_2\alpha_{q_2}\alpha_{q_3}+1)({\bf q}_2{\bf
q}_3).\nonumber
\end{eqnarray}
\section{Додаток 2}
\begin{eqnarray}
f_1=\left(q_1^2+q_2^2+q_3^2\right)\left(1-\frac{1}{\alpha_{q_1}}\right)
\left(1-\frac{1}{\alpha_{q_2}}\right)\left(1-\frac{1}{\alpha_{q_3}}\right).\nonumber
\end{eqnarray}
\begin{eqnarray}
f_1'=\frac{1}{2}\left(q_1^2+q_2^2+q_3^2\right)\sum_{1\leq
i<j\leq3}\frac{\alpha^2_{q_i}-1}{\alpha^3_{q_i}}\left(1-\frac{1}{\alpha_{q_j}}\right)\left(1-\frac{1}{\alpha_{q_k}}\right).\nonumber
\end{eqnarray}
\begin{eqnarray}
f_1''&=&\frac{1}{4^2}\left(q_1^2+q_2^2+q_3^2\right)\sum_{1\leq
i<j\leq3}\left\{-3\frac{(\alpha^2_{q_i}-1)^2}{\alpha^5_{q_i}}
\left(1-\frac{1}{\alpha_{q_j}}\right)\left(1-\frac{1}{\alpha_{q_k}}\right)\right.
+\left.2\frac{\alpha^2_{q_i}-1}{\alpha^3_{q_i}}
\frac{\alpha^2_{q_j}-1}{\alpha^3_{q_j}}\left(1-\frac{1}{\alpha_{q_k}}\right)\right\}.\nonumber
\end{eqnarray}
\begin{eqnarray}
f_2=\sum\limits_{1\leq i<j\leq3}({\mathbf q}_i{\mathbf
q}_j)(\alpha_{q_i}-1)(\alpha_{q_j}-1).\nonumber
\end{eqnarray}
\begin{eqnarray}
f_2'=\frac{1}{2N}\sum\limits_{1\leq i<j\leq3}({\mathbf
q}_i{\mathbf
q}_j)\left\{(\alpha_{q_i}-1)\frac{\alpha^2_{q_j}-1}{\alpha_{q_j}}+(\alpha_{q_j}-1)
\frac{\alpha^2_{q_i}-1}{\alpha_{q_i}}\right\}.\nonumber
\end{eqnarray}
\begin{eqnarray}
f_2''&=&\frac{1}{4N^2}\sum\limits_{1\leq i<j\leq3}({\mathbf
q}_i{\mathbf
q}_j)\left\{2\frac{\alpha^2_{q_i}-1}{\alpha_{q_i}}\frac{\alpha^2_{q_j}-1}{\alpha_{q_j}}
-(\alpha_{q_i}-1)\frac{(\alpha^2_{q_j}-1)^2}{\alpha^3_{q_j}}\right.
-\left.(\alpha_{q_j}-1)\frac{(\alpha^2_{q_i}-1)^2}{\alpha^3_{q_i}}\right\}.\nonumber
\end{eqnarray}
\begin{eqnarray}
f_3=\alpha_{q_1}\alpha_{q_2}\alpha_{q_3}\sum\limits_{j=1}^3q_j^2\alpha_{q_j}.\nonumber
\end{eqnarray}
\begin{eqnarray}
f_3'=\frac{1}{2N}\sum\limits_{j=1}^3q_j^2\left\{2\alpha_{q_1}\alpha_{q_2}\alpha_{q_3}\frac{\alpha^2_{q_j}-1}{\alpha_{q_j}}+\alpha^2_{q_j}\alpha_{q_k}
\frac{\alpha^2_{q_i}-1}{\alpha_{q_i}}+\alpha^2_{q_j}\alpha_{q_i}
\frac{\alpha^2_{q_k}-1}{\alpha_{q_k}}\right\}.\nonumber
\end{eqnarray}
\begin{eqnarray}
&&f_3''=\frac{1}{4N^2}\sum\limits_{j=1}^3q_j^2\left\{-\alpha^2_{q_j}\alpha_{q_k}
\frac{(\alpha^2_{q_i}-1)^2}{\alpha^3_{q_i}}+\alpha^2_{q_j}\alpha_{q_i}
\frac{(\alpha^2_{q_k}-1)^2}{\alpha^3_{q_k}}\right.\nonumber\\
&+&\left.4\alpha_{q_i}\alpha_{q_j}
\frac{\alpha^2_{q_j}-1}{\alpha_{q_j}}\frac{\alpha^2_{q_k}-1}{\alpha_{q_k}}+4\alpha_{q_j}\alpha_{q_k}
\frac{\alpha^2_{q_j}-1}{\alpha_{q_j}}\frac{\alpha^2_{q_i}-1}{\alpha_{q_i}}+2\alpha^2_{q_j}
\frac{\alpha^2_{q_i}-1}{\alpha_{q_i}}\frac{\alpha^2_{q_k}-1}{\alpha_{q_k}}\right\}.\nonumber
\end{eqnarray}
\begin{eqnarray}
f_4=\sum_{1\leq i<j\leq3}{({\mathbf q}_i{\mathbf
q}_j)}\left(1-\frac{1}{\alpha_{q_i}}\right)
\left(1-\frac{1}{\alpha_{q_j}}\right).\nonumber
\end{eqnarray}
\begin{eqnarray}
f_4'=\frac{1}{2N}\sum\limits_{1\leq i<j\leq3}({\mathbf
q}_i{\mathbf
q}_j)\left\{\left(1-\frac{1}{\alpha_{q_i}}\right)\frac{\alpha^2_{q_j}-1}{\alpha^3_{q_j}}+\left(1-\frac{1}{\alpha_{q_j}}\right)
\frac{\alpha^2_{q_i}-1}{\alpha^3_{q_i}}\right\}.\nonumber
\end{eqnarray}
\begin{eqnarray}
&&f_4''=\frac{1}{4N^2}\sum\limits_{1\leq i<j\leq3}({\mathbf
q}_i{\mathbf
q}_j)\left\{2\frac{\alpha^2_{q_i}-1}{\alpha^3_{q_i}}\frac{\alpha^2_{q_j}-1}{\alpha^3_{q_j}}
-3\left(1-\frac{1}{\alpha_{q_i}}\right)\frac{(\alpha^2_{q_j}-1)^2}{\alpha^5_{q_j}}\right.
-\left.3\left(1-\frac{1}{\alpha_{q_j}}\right)\frac{(\alpha^2_{q_i}-1)^2}{\alpha^5_{q_i}}\right\}.\nonumber
\end{eqnarray}


\begin{thebibliography}{99}

\bibitem{B1947} N. N. Bogoliubov, J. Phys. (USSR) 9, 23 (1947). 
\bibitem{BZ1955} Н. Н. Боголюбов, Д. Н. Зубарєв, Журн. эксп. и теор. физ. {\bf 28},129 (1955).
\bibitem{BS1957} K.A. Brueckner, K.Sawada, Phys.Rev. {\bf 106},1117 (1957).
\bibitem{VakUhn79_VHU79} И. А. Вакарчук, И. Р. Юхновський, Теор. мат. физ. {\bf 40},100 (1979); 
И. А. Вакарчук, О. Л. Гонопольський, И. Р. Юхновський, Теор. мат. физ. {\bf 41},77 (1979).
\bibitem{V1985_89_90} И.А. Вакарчук, Теор.мат.физ. {\bf 65 }, 285 (1985); {\bf 80},439 (1989); {\bf 82}, 438 (1990). 
\bibitem{V1988_VH} И.А. Вакарчук. П.А.Глушак,  Теор.мат.физ. {\bf 75}, 101 (1988); І. О. Вакарчук, П. А. Глушак, Укр. фіз. журн. {\bf
41}, 569 (1996).
\bibitem{Hlushak} П. А. Глушак. Исследование равновесных
свойств сверхтекучево гелия-4 при низьких температурах.
Кандидатская диссертация. Львов (1992).
\bibitem{Vak_Bab_Rov} I. O. Vakarchuk, V. V. Babin, A. A. Rovenchak, J. Phys. Stud. {\bf 4}, 16 (2000);
\bibitem{Vak_Rov12}   I. O. Vakarchuk, A. A. Rovenchak, J. Phys. Stud. {\bf 5}, 126 (2001); {\bf 4}, 431 (2001).
\bibitem{VPR2007} І. О. Вакарчук, Р. О. Притула, А. А. Ровенчак, Журн. фіз. досл. {\bf 11}, 259 (2007).
\bibitem{expe1} W. L. McMillan, Phys. Rev. A {\bf 138}, 442
(1965).
\bibitem{expe2} D. Schiff, L. Verlet, Phys. Rev. {\bf 160}, 208
(1967).
\bibitem{expe3} V. F. Sears, Phys. Rev. B {\bf 28}, 5109 (1983).
\bibitem{expe4} V. F. Sears, R. D. McCarty, D. G. Friend, NIST Technical Note 1334 (1998).
\bibitem{Temperly} Г. Темперли, Дж. Роулисона, Дж. Рашбрука, {\it Физика простих жидкостей} (Мир, Москва, 1971).
\bibitem{Krokston} К. Крокстон, {\it Физика жидкого состояния} (Мир, Москва, 1978).



\bibitem{MaEd}  H. J. Maris, D. O. Edwards, J. Low. Temp. Phys, {\bf 129}, 1 (2002). 
\bibitem{LRCG}  T. Lindenau, M. L. Ristig, J. W. Clark, K. A. Gernoth, J. Low. Temp. Phys, {\bf 129}, 143 (2002). 
\bibitem{SBBLD} D. A. Sergatskov,  A. V. Babkin, S. T. P. Boyd, R. A. M. Lee,  R. V. Duncan, 
J. Low. Temp. Phys, {\bf 134}, 517 (2004).

\bibitem{Mosameh} S. M. Mosameh, A. S. Sandouqa, H. B. Ghassib, B. R. Joudeh, J. Low. Temp. Phys, {\bf 175}, 523 (2014).

\bibitem{KrCh} B. Krishnamachari and G. V. Chester, Phys. Rev. B 61, 9677 (2000).  
\bibitem{CaMi} Fr\'ed\'eric  Caupin, Tomoki Minoguchi, J. Low. Temp. Phys, {\bf 134}, 181 (2004).
\bibitem{Kim} E.Kim,  M. H. W. Chan, Science {\bf 305},1941, (2004); Phys. Rev. Lett. {\bf 97},115302, (2006).
\bibitem{HRV} R. B. Hallock, M. W. Ray, Y. Vekhov, J. Low. Temp. Phys, {\bf 169}, 264 (2012).
\bibitem{Chan} M. H. W. Chan, R. B. Hallock, L. Reatto,  J. Low. Temp. Phys, {\bf 172}, 317 (2013).


\bibitem{CCK} C. E. Campbell, B. E. Clements, E. Krotscheck, and M. Saarela, Phys. Rev. B, {\bf 55} 3769 (1997) 
\bibitem{DKFG} Manuel Diaz-Avila,  Mark O. Kimball, Francis M. Gasparini, J. Low. Temp. Phys, {\bf 134}, 613 (2004). 
\bibitem{ADM}  R. H. Anderson, David Z. Li, M. D. Miller, J. Low. Temp. Phys, {\bf 169}, 291 (2012).


\bibitem{FoMo} R. Folk, G. Moser, J. Low. Temp. Phys, {\bf 150}, 689 (2008); 
\bibitem{ChBr} Gunaranjan Chaudhry, J. G. Brisson, J. Low. Temp. Phys, {\bf 155}, 235 (2009); {\bf 158}, 806 (2010).


\bibitem{LCS}  J. A. Lipa, M. Coleman,  D. A. Stricker, J. Low. Temp. Phys, {\bf 124}, 443 (2001).
\bibitem{SDG} Ali Shams, J. L. DuBois, H. R. Glyde, J. Low. Temp. Phys, {\bf 145}, 357 (2006).
\bibitem{GoBo}  M. C. Gordillo,  J. Boronat,  J. Low. Temp. Phys., {\bf 171}, 606 (2013).


\bibitem {BGC} J. Boronat, M. C. Gordillo, J. Casulleras, J. Low. Temp. Phys, {\bf 126}, 199 (2002). 
\bibitem {BMK} I. Be\v{s}li\'c, L. Vranje\v{s} Marki\'c, S. Kili\'c,  J. Low. Temp. Phys, {\bf 143}, 257 (2006). 
\bibitem {Vitiello} S. A. Vitiello,  J. Low. Temp. Phys, {\bf 162}, 154 (2011). 
 
 
\bibitem{Ceperley} D. M. Ceperley, Rev. Mod. Phys. {\bf 67}, 279 (1995).
\bibitem{BSSC} J. Boronat, K. Sakkos, E. Sola, J. Casulleras, J. Low. Temp. Phys, {\bf 148}, 845 (2007).
\bibitem{Robo} R. Rota, J. Boronat, J. Low. Temp. Phys, {\bf 162}, 146 (2011).


\bibitem{GFMC}  M. H. Kalos, Phys. Rev. {\bf 128}, 1791 (1962).

\bibitem{MoBo2}   S. Moroni,  M. Boninsegni, J. Low. Temp. Phys, {\bf 136}, 129 (2004).

\bibitem{SL_KR}  J. C. Slater, J. G. Kirkwood, Phys. Rev. {\bf 37}, 682 (1931).

\bibitem{MiSc} Kunimasa Miyazaki and I. M. de Schepper, Phys. Rev. E, {\bf 63}, 060201(R) (2001).
\bibitem{EKZ}  J. Egger, E. Krotscheck,  R. E. Zillich, J. Low. Temp. Phys, {\bf 165}, 275 (2011).


\bibitem{AZ_SL1} R. A. Aziz,  M. J. Slaman, J. Chem.Phys. {\bf 94}, 8047 (1991).
\bibitem{AZ_SL2}  R. A. Aziz,  M. J. Slaman, A. Koide, A. R. Allnatt, W. J.
Meath, Mol. Phys. {\bf 77}, 321 (1992).
\bibitem{Br_Cs} J. Boronat, J. Casulleras, Phys. Rev. B. {\bf 49}, 8920(1994).
\bibitem{Ak_Tl} B. M. Axilrod, E. Teller, J. Chem. Phys. {\bf 11}, 299 (1943).
\bibitem{Pr_Dk} C. A. Parish, C. E. Dykstra, J. Chem. Phys. {\bf 9}, 7618 (1994).
\bibitem{Rov_dys} A. А. Ровенчак, Самоузгоджений розрахунок міжатомних
потенціалів та термодинамічних функцій гелію-4 в надплинній та
нормальній фазах. Кандидатська дисертація. Львів (2003).
\bibitem{VakHryh1} І. О. Вакарчук, О. І. Григорчак, Журн. фіз. досл. {\bf 3}, 3005 (2009).
\bibitem{VakHryh2} І. О. Вакарчук, О. І. Григорчак, Вісник Львівського університету. Серія фізична. {\bf 46}, 3 (2011).
\bibitem{VakHryh3} І. О. Вакарчук, О. І. Григорчак,  arXiv:1506.03707 (2015).
\bibitem{HP2014}  І. О. Вакарчук, О. І. Григорчак, В. С. Пастухов, Р. О. Притула,  arXiv:1506.03317 (2015).
\bibitem{Vak2004} I. O. Vakarchuk,  J. Phys. Stud. {\bf 8}, 223 (2004).
\bibitem{Vstup} І. Вакарчук, {\it Вступ до проблеми багатьох тіл} (Львівський національний університет імені Івана Франка, Львів, 1999).
\bibitem{DonBar} R. J. Donnelly, C. F. Barenghi, J. Phys. Chem. Ref. Data, {\bf 27}, 6 (1998).
\end{thebibliography}
\end{document}